\documentclass[useAMS,usenatbib]{mn2e}
\usepackage{graphics,epsfig,psfig}
\usepackage[normalem]{ulem}
\usepackage[dvipsnames]{color}
\usepackage[]{inputenc,amssymb}
\usepackage{amsmath}
\usepackage{color}
\usepackage{amssymb}
\usepackage{cancel}

\usepackage[toc,page]{appendix}

\def \be{\begin{equation}}
\def \ee{\end{equation}}
\def \bea{\begin{eqnarray}}
\def \eea{\end{eqnarray}}
\def\etal{{et al.}}

\def \cl{C$_{\ell}$ }
\def \fg{$f_{\rm gas}$ }
\def \ftemp{$f_{\rm Temp}$ }
\def \fratio{$f_{\rm tratio}$ }

\def \pra{{\sf Prior-1} }
\def \prb{{\sf Prior-2} }
\def \prc{{\sf Prior-3} }
\def \spt{{\sf SPT-like} }
\def \cv{{\sf CV1000} }

\title[CMB distortion from circumgalactic gas] 
{CMB distortion from circumgalactic gas} 

\author[Singh, Nath, Majumdar \& Silk]
{Priyanka Singh$^{1}$ \thanks{priyankas@rri.res.in},
 Biman B. Nath$^1$, 
 Subhabrata Majumdar$^{2}$ 
 \& Joseph Silk$^{3,4}$
 \\
$^1$ Raman Research Institute, Sadashivanagar, Bangalore, India, 560080 \\
$^2$ Tata Institute of Fundamental Research, Mumbai, India, 400005\\
$^3$ Institut d'Astrophysique, 75014Paris, France\\
$^4$ The John Hopkins University, Baltimore MD 20218, USA
}

\begin{document}


\maketitle

\label{firstpage}

\begin{abstract}
We study the Sunyaev-Zel'dovich (SZ) distortion of the cosmic microwave background radiation (CMBR) 
from extensive circumgalactic gas (CGM) in massive galactic halos. Recent observations have shown that galactic halos contain
a large amount of X-ray emitting gas at the virial temperature, as well as a significant amount of warm OVI absorbing gas. We 
consider the SZ distortion from the hot gas in those galactic halos in which the gas cooling time is longer than the halo
destruction time scale. We show 
that the SZ distortion signal from the hot gas in these galactic halos at redshifts $z\approx 1\hbox{--}8$ can be significant at small
angular scales ($\ell\sim 10^4$), and dominate over the signal from galaxy clusters. The estimated SZ signal for most massive galaxies (halo mass $\ge 10^{12.5}$ M$_\odot$) is consistent with the marginal detection by {\it Planck} at these mass scales.
 We also consider the SZ effect from warm circumgalactic gas. The integrated
Compton distortion from the warm OVI absorbing gas is estimated to be $y\sim 10^{-8}$, which could potentially be detected by 
experiments planned for the near future. Finally, we study the detectability of the SZ signal from circumgalactic gas in two
types of surveys, a simple extension of the SPT survey and a more futuristic cosmic variance-limited survey. We find that these surveys
can easily detect the kSZ  signal from  CGM. With the help of a Fisher Matrix analysis, we find that it will be possible for these surveys to constrain
the gas fraction in CGM, after marginalizing over cosmological parameters, to $\le 33$\%, in case of no redshift evolution of the gas fraction.
\end{abstract}

\begin{keywords} 
Galaxies: halos --- evolution; cosmology: cosmic microwave background
\end{keywords}

\section{Introduction}
The standard scenario of galaxy formation predicts that baryonic gas falls into dark matter potentials and 
gets heated to the virial temperature \citep{silk77, white78, white91}. This gas then cools radiatively, and if the
temperature is low enough ($T \le 10^6$ K) for significant radiation loss, then most of the galactic halo gas
drops to low temperature and no accretion shock develops in the halo \citep{birnboim03}. In the case of low mass galaxies,
most of the accretion takes place through the infall of cold material from the intergalactic medium (IGM). However,
in  massive galaxies, the hot halo gas cools slowly and should remain warm/hot for a considerable period of time. This halo gas, if present,
could potentially contain a large fraction of the baryons in the universe which is unaccounted for by collapsed
gas and stars in galaxies, and could explain the missing baryon problem \citep{fukugita98, anderson10}. 

Although numerical simulations have
shown that disc galaxies should be embedded in a hot gaseous halo, this gas has been difficult
to nail down observationally because of faintness of the X-ray emission \citep{benson00, rassmussen09, crain10}. 
Recent observations have finally discovered this hot coronal gas extended over a large region around massive
spiral galaxies \citep{anderson11, dai12, anderson13, bogdan13a, bogdan13b, anderson14, walker14}. The typical densities at galactocentric
distances of $\ge 100$ kpc is inferred to be a few times $10^{-4}$ cm$^{-3}$ (e.g., \cite{bogdan13b}), at a temperatures of $\sim 0.5$ keV. The amount of material implied in this extended region is unlikely to come from the star formation process, as shown by \cite{bogdan13a}. 
An extended region of circumgalactic medium (CGM) has also been observed through OVI absorption lines
around massive galaxies at $z\le 1$ \citep{tumlinson11}, although these observations probe clouds at $T \sim 10^{5.5}$ K.

At the same time, the presence of hot halo gas around the Milky Way galaxy has been inferred via ram pressure arguments from the
motion of satellite galaxies \citep{putman09, putman12, gatto13}. These observations suggest that the density profile of the hot coronal gas in our Galaxy is rather flat out to large radius, with $n \sim 10^{-3.5}$ cm$^{-3}$. Theoretically, one can understand this profile from simple modelling of hot, high entropy gas in hydrostatic equilibrium \citep{maller04,  sharma12, fang13}. While in galaxy clusters, the high entropy of the diffuse gas produces a core, for massive galaxies (with implied potential wells shallower than in galaxy clusters), the core size is relatively large and extends to almost the virial radius.

One of the implications of this hot coronal gas in the halos of massive galaxies is the SZ distortion of the CMBR \citep{planck13}. 
The average $y$ distortion of the CMBR from massive galaxies is likely to be small. However, the anisotropy power spectrum could have a substantial contribution from the hot gas in galactic halos.
The SZ distortion from galaxy clusters have been computed with the observed density and temperature profiles of the X-ray emitting gas, or the combined pressure profile (e.g, \cite{majumdar01, komatsu02, efstathiou12}). In the case of the galactic halos, because of the expected flat density profile, the resulting $y$-distortion could be larger than that of galaxy clusters for angular scales that correspond to the virial radii of massive galaxies, i.e., $\ell\sim 10^4$. These angular scales are being probed now and therefore the contribution to the  SZ signal from galactic halos is important.
In this paper, we calculate the angular power spectrum from both the thermal and kinetic SZ effects, if a fraction $f\sim 0.11$ of the total baryonic content of massive galaxies is in the form of hot or ionized halo gas.

Although such a fraction of gas has been estimated from the observations of NGC 1961 and NGC 6753 \citep{bogdan13a}, it remains uncertain whether it is a representative value, or whether it can be as low as $0.05$. Recent studies of absorption from halo gas along the lines of sight to background quasars show that roughly half of the missing baryons is contained in the halo as  warm (at $\sim 10^4$ and $\sim 10^{5.5}$ K) components. We also discuss the possible SZ signatures from this cool-warm gas in galactic halos.

\section{Sunyaev-Zel'dovich distortion from hot galactic halo gas}
For simplicity, we assume that galactic halos contain a constant fraction of the total halo mass, independently of the galaxy mass. If we consider the total baryon fraction $\Omega_b/\Omega_m \sim 0.16$, and the fraction of the
total mass that is likely to be in the disc, which is predicted to be $\sim 0.05$ \citep{mo98, moster10, leau10, dutton10}, then one can assume
a fraction $f_{\rm gas}\,=\,0.11$ of the total halo mass to be spread throughout the halo. We also assume it to be uniform in density, with 
a temperature given by the virial temperature of the halo. The uncertainties in gas fraction and temperature are explored later in Section \ref{sec-fisher}.
The cosmological parameters needed for our calculatiosn are taken from the recent Planck results (Table 2 of \cite{planck13a}). 

\begin{figure}
\begin{center}
  \centerline{\epsfxsize=0.5\textwidth\epsfbox{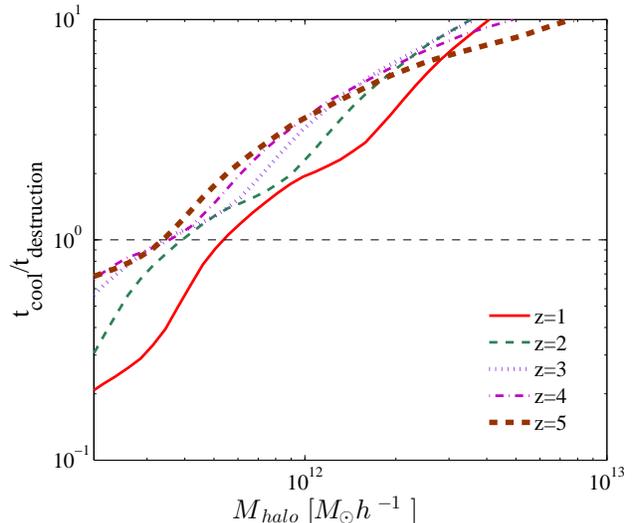}}
{\vskip-3mm}
 \caption {The ratio of cooling time to destruction time scale of halos is shown as a function
 of halo mass collapsing at $z = 1$ (red solid line), $z = 2$ (thin green dashed line), $z = 3$ (blue  dotted line), $z = 4$
 (magenta dot-dashed line) and $z = 5$ (thick brown dashed line). }
 \label{cool}
\end{center}
\end{figure}

\subsection{Thermal Sunyaev-Zel'dovich effect}
When CMBR photons are inverse Compton scattered by high energy electrons,  the CMB spectrum is distorted giving rise to the thermal Sunyaev-Zel'dovich effect (tSZ). This effect is represented in terms of the Compton y-parameter defined as $
y= (k_b T_e n_e \sigma_TL)/(m_e c^2)$
where $\sigma_T$ is the Thomson scattering cross section, $T_e$ is the temperature ($T_e>>T_\gamma$) and $n_e$ is the electron density of the medium, considered to be uniform here, and $\ell$ is the distance traversed by the photons through the medium. The profile of $y$ can
be written in terms of the impact parameter $w$, or the angle $\theta=w/D_A$ (where $D_A$ is the angular diameter distance) as
\begin{eqnarray}
 y(w)&=&\frac{2k_bT_vn_e\sigma_T}{m_ec^2}\sqrt{R_v^2-w^2} \,,\nonumber\\
 y(\theta)&=&\frac{2k_bT_vn_e\sigma_TR_v}{m_ec^2}\sqrt{1-\frac{D_A^2\theta^2}{R_v^2}}
 \label{eqn-ytheta}.
\end{eqnarray}
Here the electron density $ n_e =\frac{\rho_{gas}}{\mu_e m_p}$ of the hot gas is determined by the requirement that the total hot gas mass within the virial radius 
is a fraction $f_g=0.11$ of the halo mass. 
The virial radius of a halo of mass M collapsing at redshift z is given by
\begin{equation}
 R_{vir}=0.784 \Bigl( \frac{M}{10^8 h^{-1}}\Bigr)^{1/3} \Bigl( \frac{\Omega_M }{\Omega_M(z)} \frac{\triangle(z)}{18 \pi^2}\Bigr)^{-1/3} \Bigl(\frac{1+z}{10} \Bigr)^{-1} h^{-1} {\rm kpc}
\end{equation}
where $\Omega_M(z)=\Omega_M(1+z)^3/E^2(z)$, the critical overdensity $\triangle(z)=18 \pi^2+82d-39d^2$ and $d=\Omega_M(z)-1$.\\
Later, we will also discuss the effect of varying $f_g$, including its possible redshift evolution.
The temperature $T_e$ corresponds to the virial temperature of the halo. We discuss in \S 3.2 below
the appropriate mass and redshift range of galactic halos in which the gas likely remains hot.


\subsection{Kinetic Sunyaev-Zel'dovich effect}
If the scattering medium has bulk velocity with respect to the CMB frame, the CMBR is anisotropic in the rest frame of the scattering medium. The scattering makes the CMBR isotropic
in the rest frame of the scattering medium, resulting in the distortion of the CMB spectrum with respect to the observer and giving rise to the  kinetic Sunyaev-Zel'dovich
effect (kSZ). The kSZ effect is proportional to the line of sight peculiar velocity and optical depth of the scattering medium. In the non-relativistic limit,
the Compton y-parameter for the kSZ effect is defined as $y= (v_{\rm los} n_e \sigma_T L)/c$
where $v_{\rm los}$ is the line-of-sight peculiar velocity of the scattering medium. The tSZ effect and the kSZ effect have different frequency dependences which makes them easily separable
with good multi-frequency data. 
In contrast to the  tSZ effect, the spectral shape of the CMB is unchanged by the  kSZ effect. In the Rayleigh-Jeans limit, the ratio of the change in CMB temperature
caused by these two effects is :
\begin{eqnarray}
 \frac{\vartriangle T_{\rm kin}}{\vartriangle T_{\rm th}} &\approx& \frac{1}{2} \frac{v_{\rm los}}{c} \Bigl (\frac{k_b T_e}{m_e c^2}\Bigr )^{-1} \,,\nonumber\\
 &\approx& 0.09 \Bigl (\frac{v_{\rm los}}{1000 \,{\rm km} \, {\rm s}^{-1}}\Bigr ) \Bigl (\frac{k_b T_e}{10 \, {\rm kev}}\Bigr )^{-1}
 \label{eq:ksztotsz}
\end{eqnarray}
For galaxy clusters, $k_b T_e \sim 10$ keV and $v_{\rm los}\sim$ few hundred km/sec which makes tSZ $\gg$ kSZ . But for the  case of galaxies with virial temperature
$T_e\sim 10^6$ K, hence $k_b T_e \sim 0.1$ keV, thus making kSZ $>$ tSZ.

\section{The SZ Power Spectrum}
The SZ power spectrum arises by summing over the contributions from {\it all} the halos that would distort the  CMB convolved with the template distortion for the halos as a function of mass and redshift; the distribution of the halos can be approximated by fits to outputs from N-body simulations. However, not all dark matter halos identified in the simulations would contribute to the SZ \cl and one has to use {\it only} those galactic halos where the gas has not cooled substantially; this is discussed in detail in Section \ref{sec-halocool}. 


\begin{figure} 
\centerline{\epsfxsize=0.55\textwidth\epsfbox{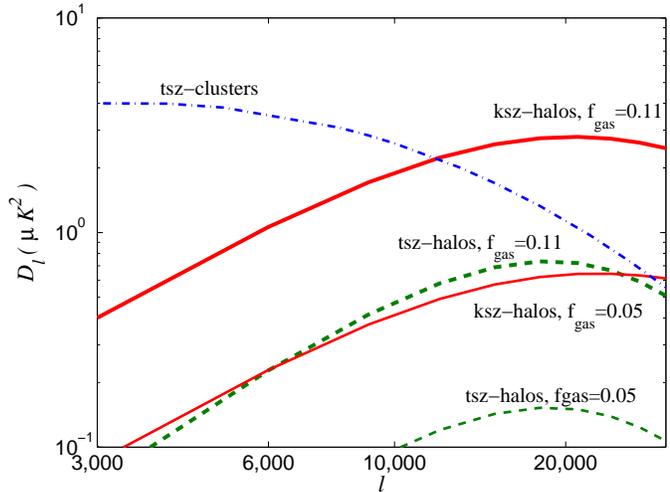} }
{\vskip-5mm}
 \caption {Angular power spectrum of CMBR at 150 GHz over a larger range of $\ell$, for tSZ  (green dashed line) and kSZ (red solid line) from galactic halos,
 compared with tSZ from clusters (blue dot-dashed line). Here the thick and thin lines correspond to $f_{gas}=0.11$
 and $f_{gas}=0.05$ respectively.
 }
 \label{largel}
\end{figure}


\subsection{Thermal SZ \cl}
The thermal SZ template for contribution by a galactic halo is given by the angular Fourier transform of $y(\theta)$ (see Equation \ref{eqn-ytheta}) and is given by 
\begin{eqnarray}
 y_l&\approx&2\pi\int_0 ^\pi\theta y(\theta)J_o[(l+1/2)\theta]d\theta \,, \nonumber\\
&=&\frac{4\pi k_b\sigma_TR_v}{m_ec^2}\int_0 ^\pi \theta T_v n_e \sqrt{1-\frac{D_A^2\theta^2}{R_v^2}}J_o \Bigl [ (l+1/2)\theta \Bigr ] d\theta \nonumber\\
&=& \frac{8 k_bT_vn_e\sigma_T R_v^{3/2}}{m_ec^2 D_A^{1/2}}(\frac{\pi}{2l+1})^{3/2}J_{3/2}\Bigl [ (l+1/2)\frac{R_v}{D_A}\Bigr ]\,.
\end{eqnarray}
The last equality follows from \cite{gradshteyn80}.

The angular power spectrum due to the tSZ effect by hot diffuse gas in galactic halos is given by
\begin{equation}
 C_l=g^2(x)C_l^{yy}
\end{equation}
Where $g(x)=x \coth(x/2)-4 $ and $C_l^{yy}$ is frequency independent power spectrum. 
\begin{equation}
 C_l^{yy}=C_l^{yy(P)}+C_l^{yy(C)}
\end{equation}
where $C_l^{yy(P)}$ is the Poisson term and $C_l^{yy(C)}$ is clustering or correlation term. These two terms can be written as
\citep{komatsu99}
\begin{eqnarray}
 C_l^{yy(P)}&=&\int_0 ^{z_{\rm max}} dz \frac{dV}{dz}\int_{M_{min}} ^{M_{\rm max}} dM\frac{dn(M,z)}{dM} |y_l(M,z)|^2 \, \nonumber\\
 C_l^{yy(c)}&=&\int_0 ^{z_{\rm max}} dz \frac{dV}{dz} P_m(k=\frac{l}{r(z)},z)\nonumber\\
 &&\times \Bigl[\int_{M_{min}} ^{M_{\rm max}} dM\frac{dn(M,z)}{dM} b(M,z) y_l(M,z)\Bigr]^2 \,.
 \label{eqn-clyy}
\end{eqnarray}
Here $r(z)=(1+z)D_A$ is the comoving distance, $\frac{dV}{dz}$ is differential comoving volume per steradian, $P_m(k,z)$ is matter power spectrum, $b(M,z)$  is the linear
bias factor,  and $\frac{dn(M,z)}{dM}$ is the differential mass function. Here we have used the Sheth-Tormen mass function 
\begin{eqnarray}
  \frac{dn}{dM}dM&=&A \sqrt{{2\alpha \nu^2 \over \pi}} \frac{\rho_m}{M^2}e^{-\alpha \nu^2}
  \Bigl [-\frac{d \log\sigma}{d \log M}\Bigr ]\nonumber\\
  && \times \Bigl [1+\Bigl (\alpha \nu^2 \Bigr )^{-p} \Bigr ]dM \,,
  \label{eqn-STmassfn}
\end{eqnarray}

where $A=0.322184$, $\alpha=0.707$ and $p=0.3$ \citep{sheth01}. We have used the bias factor from 
\cite{jing99},
\begin{equation}
   b(M,z)=\Bigl (1+\frac{0.5}{\nu^4}\Bigr )^{(0.06-0.02n)} \Bigl (1+\frac{\nu^2-1}{\delta_c} \Bigr )
\end{equation}
with $\nu=\frac{\delta_c}{D_g(z)\sigma(M)}$, where $D_g(z)$ is the growth factor, n is the index of primordial power
spectrum , $\delta_c=1.68$ is the critical overdensity and $\sigma(M)$ is the present day smoothed (with top hat filter) variance.

\begin{figure}
\begin{center}
\includegraphics[width=8.8cm,angle=0 ]{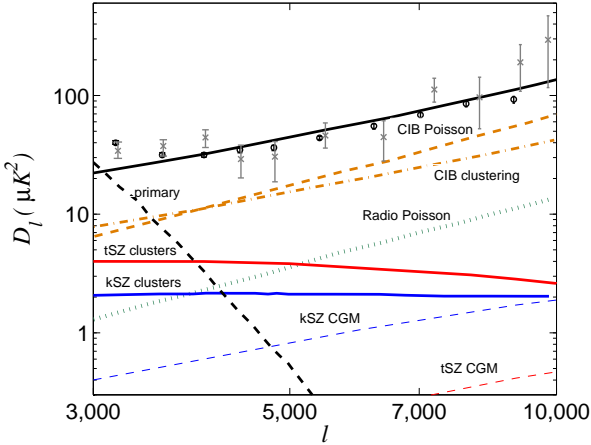}
\caption {Angular power spectrum of CMBR at 150 GHz from different processes and compared with
 data points from ACT (grey bars) and SPT (black bars). The thermal SZ from galactic halos calculated here
is shown in red (thick solid line), and the kSZ from galactic halos is shown in blue (thick solid line). Thermal SZ from galaxy clusters \citep{efstathiou12} is shown in red (thin solid line)
, the kinetic SZ from galaxy clusters is shown in blue (thin solid line), the radio poisson
and CIB poisson signals in green dotted and brown dashed lines, respectively, and CIB clustering signal is shown in brown dot-dashed line. The lensed primary signal is shown in black and the total
signal 
is shown by grey line.}
\label{spt}
\end{center}
\end{figure}


\subsection{Mass and redshift range}
\label{sec-halocool}
As mentioned earlier, not all the galactic halos given by the ST mass function (i.e Equation \ref{eqn-STmassfn}) will contribute to the SZ \cl. For a realistic estimate of the CMB distortion from circum-galactic gas in galaxies, we need to
use only those galactic halos in which the hot halo gas does not cool substantially, so that the hot gas persists for a considerable period of time and can contribute to the anisotropy. The cooling time of the gas is defined
as $t_{\rm cool}=1.5 nkT/(n_e^2 \Lambda(T))$, where $n$ is the particle density ($\sim \frac{\mu_e}{\mu} n_e$), $\mu$ is mean molecular weight of the gas, $\nu_e$ is the mean molecular weight per free electron and $\Lambda(T)$ is the cooling function. We assume the galactic halo gas to be of metallicity $0.1$ Z$_\odot$, 
and use the cooling function from \citep{dopita93}. 

This cooling time should be compared with a time scale corresponding to the destruction of these galactic halos in the merger or accretion processes, which would lead to the formation of larger halos. Every merging event leads to heating of the halo gas back to the virial temperature. It is therefore reasonable to assume that the halo gas would remain hot at the virial temperature if the cooling time is longer than the time corresponding to the destruction of halos.

We have used an excursion set approach to calculate the destruction time \citep{lacey93, lacey94}. 
For Press-Schechter mass function the destruction
time for a galactic halo of mass M at time t is
\begin{eqnarray}
 t_{dest}(M,t)&=&[\phi(M,t)]^{-1}\,,\nonumber\\
 &=& \Bigl [\int _{M(1+\epsilon)} ^\infty \tilde{Q}(M,M_1;t)dM_1 \Bigr ]^{-1}
\end{eqnarray}
Where $\tilde{Q}(M,M_1;t)$ is the probability that an object of mass M grows into an object of mass $M_1$ per unit time through merger or 
accretion at time t. 
\begin{eqnarray}
 \tilde{Q}(M,M_1;t)dM_1&=&\sqrt{\frac{2\sigma^2(M_1)}{\pi}}\Bigl [\frac{\sigma^2(M)}{\sigma^2(M_1)(\sigma^2(M)-\sigma^2(M_1))}\Bigr ]^\frac{3}{2} \nonumber\\
 &&\times  \Bigl |\frac{d\delta}{dt} \Bigr |{\rm exp} \Bigl [-\frac{\delta^2(\sigma^2(M)-\sigma^2(M_1))}{2\sigma^2(M)\sigma^2(M_1)} \Bigr]\nonumber\\
 &&\times \Bigl |\frac{d\sigma(M_1)}{dM_1} \Bigr |dM_1
\end{eqnarray}
Here we have used $\epsilon=0.1$. For the mass range considered, the destruction time for Sheth-Tormen mass function and Press-Schechter mass function
give similar results (\cite{mitra11}). For simplicity we have used the Press-Schechter mass function to calculate the destruction time.

\begin{figure}
 \centerline{\epsfxsize=0.5\textwidth\epsfbox{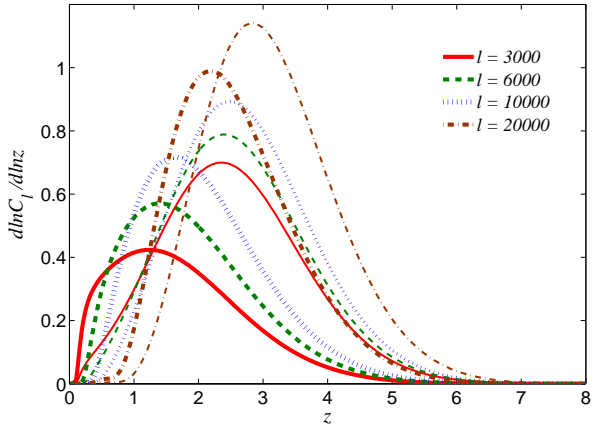}}
{\vskip-3mm}
 \caption {Redshift distribution of thermal and kinetic SZ effects. tSZ cases are shown with thin lines and kSZ cases, with thick lines for $l=3000$ (red solid lines), 
 $l=6000$ (green dashed lines), $l=10000$ (blue dotted lines) and $l=20000$ (brown dot-dashed lines).} 
 \label{zdist}
\end{figure}

We show the ratio of the cooling time to destruction timescale as a function of mass at different redshifts in Figure \ref{cool}.
Based on this estimate, we use those galactic halos in our calculation of CMBR anisotropy for which $t_{\rm cool}/t_{\rm dest} \ge 1$, so that gas in these galactic halos cannot cool quickly.
This condition is used to determine the lower mass limit of galactic halos $M_{min}$ in Equation \ref{eqn-clyy}. We have used $M_{\rm max}=10^{13} h^{-1} M_\odot$
for the upper mass limit. For upper redshift limit of integration in Equation \ref{eqn-clyy} it is sufficient to take $z_{\rm max}=8$ (see Figure \ref{zdist}).

\subsection{Kinetic-SZ \cl}
Analogously to the  tSZ effect, the angular Fourier transform of Compton y-parameter for the kSZ effect is given by:
\begin{equation}
y_l\approx 8 \frac{v_{\rm los}}{c} \frac{n_e\sigma_T R_v^{3/2}}{D_A^{1/2}}(\frac{\pi}{2l+1})^{3/2}J_{3/2}\Bigl [ (l+1/2)\frac{R_v}{D_A}\Bigr ]\,.
\end{equation}

A crucial input into the calculation of the kinetic-SZ \cl  is the line of sight peculiar velocity the dark matter halo which depends on its mass $M$, redshift and the overdensity of the environment
$\delta$ in which the halo is present \citep{sheth01a, hamana03, bhattacharya08}. 
The probability distribution function of the line of sight velocity of a halo with mass $M$ located in a region
of overdensity $\delta$ is
\begin{equation}
 p(v_{\rm los}|M,\delta,a)=\sqrt{{3 \over 2 \pi}} \frac{1}{\sigma_v(M,a)} \exp \Bigl( -\frac{3}{2}\Bigl[ \frac{v}{\sigma_v(M,a)}\Bigr]^2 \Bigr)
\end{equation}
with the 3D velocity dispersion given by
\begin{eqnarray}
 \sigma_v(M,a)&=&[1+\delta(R_{\rm local})]^{\mu(R_{\rm local})}\sigma_p(M,a)\,,\nonumber\\
 &=&[1+\delta(R_{\rm local})]^{\mu(R_{\rm local})}aH(a)D_a \Bigl( \frac{d \ln D_a}{d \ln a}\Bigr) \nonumber\\
 &&\times\Bigl( 1-\frac{\sigma^4_0(M)}{\sigma^2_{-1}(M) \sigma^2_1(M)}\Bigr)^{1/2} \sigma_{-1}(M) \,,
\end{eqnarray}
where $\sigma_p(M,a)$ is the rms peculiar velocity at the peaks of the smoothed density field and $\sigma_j$'s are the moments of initial mass distribution
defined as
\begin{equation}
 \sigma^2_j(M)=\frac{1}{2 \pi^2} \int_0 ^\infty dk k^{(2+2j)}P(k) W^2(kR(M)) \,.
\end{equation}
Here the smoothing scale $R(M)$ is given by $\Bigl( \frac{3M}{4 \pi \rho_m}\Bigr)^{1/3}$, $W(kR)$ is the top hat filter and $\rho_m$ is the present day
mean matter density. The dependence of peculiar velocity on its environment is contained in parameters $R_{\rm local}$, $\mu(R_{\rm local})$ and $\delta(R_{\rm local})$.
These parameters are obtained by the conditions \citep{bhattacharya08}
\begin{equation}
 \mu(R_{\rm local})=0.6\frac{\sigma^2_0(R_{\rm local})}{\sigma^2_0(10 \, {\rm Mpc/h})} \,.
\end{equation}
with $\sigma_0(R_{\rm local})=0.5/\sqrt{(1+z)}$ and $\delta(R_{\rm local})=\sqrt{\sigma_0(R_{\rm local})}$.

The angular power spectrum due to kSZ effect by this hot diffuse gas is independent of frequency and is given by
\begin{equation}
 C^{yy}_l=C^{yy(P)}_l+C^{yy(C)}_l
\end{equation}
\\\\Where $C^{yy(P)}_l$ and $C^{yy(C)}_l$ are Poisson and clustering terms given by Equation \ref{eqn-clyy}.

 \begin{figure}
 \begin{center}
\includegraphics[width=8.0cm, angle=0]{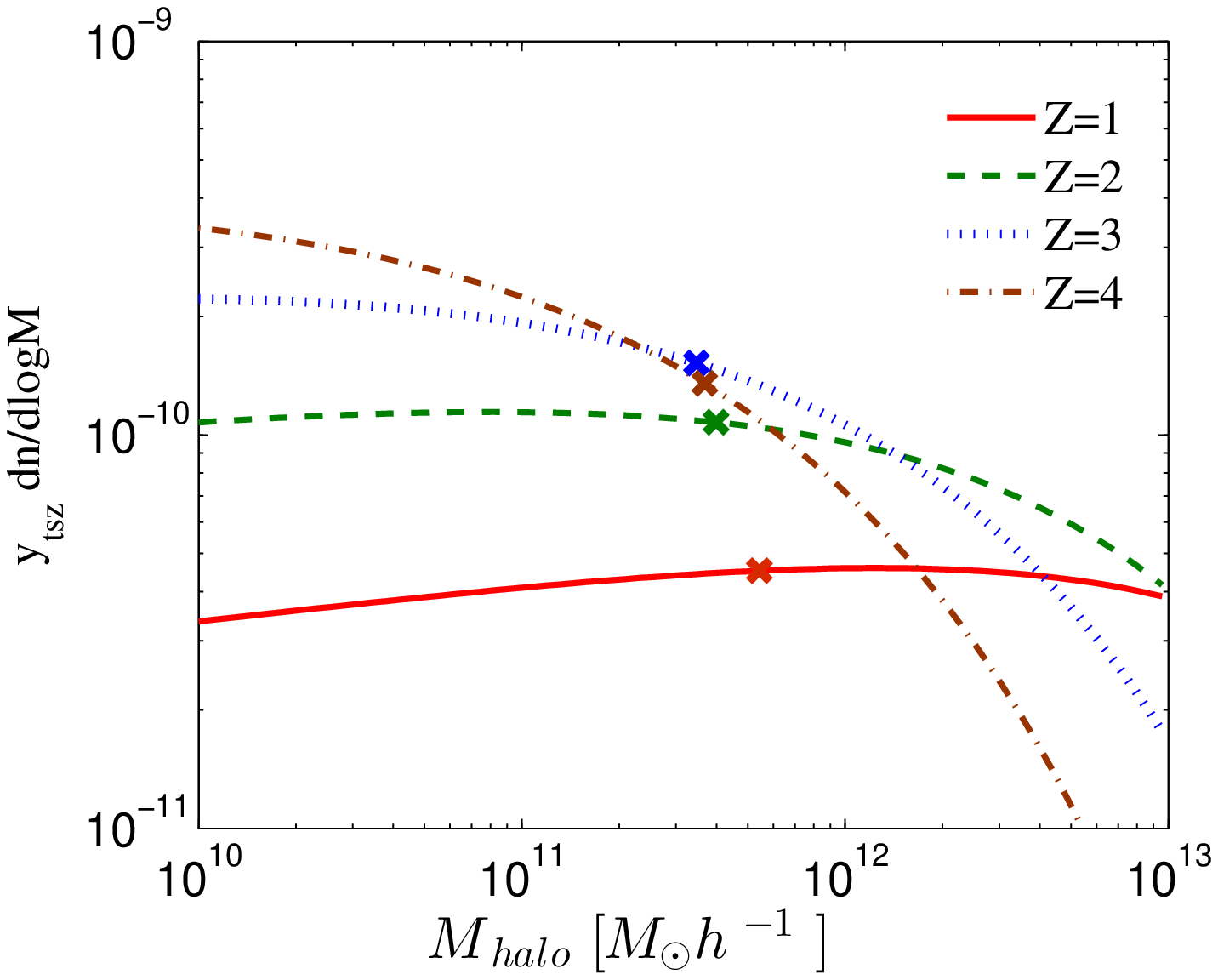}
\includegraphics[width=8.0cm, angle=0]{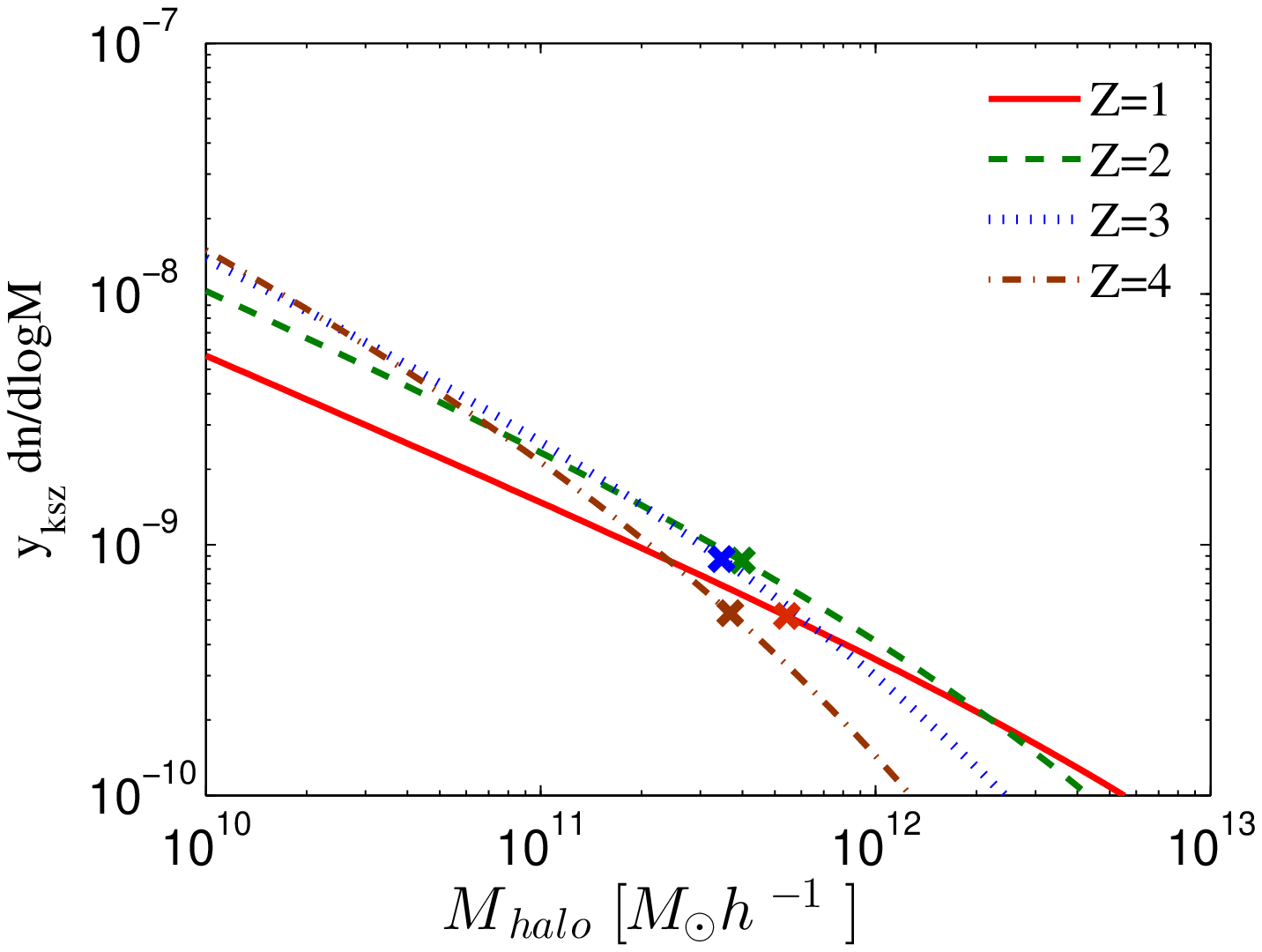}
\caption{
Moments of the mass function for tSZ (top panel) and kSZ (bottom panel), as a function of galactic halo
masses, for redshifts $z = 1$ (red solid line), $z = 2$ ( green dashed line), $z = 3$ (blue  dotted line) and $z = 4$ (brown dot-dashed line). The cross markers on each line show the lower limits of masses considered in the calculations of SZ signal based on the cooling time scale being longer than halo destruction
time scale.
}
\label{fig:moments} 
\end{center}
\end{figure}

\subsection{SZ from CGM -vs- SZ  from ICM}
We plot the multipole dependence of both thermal and kinetic SZ \cl from the CGM,  in Figure \ref{largel},  in term of the parameter  $D_l=\frac{l(l+1)}{2\pi} C_l \overline{T^2}_{CMB}$ where $\overline{T}_{CMB}$ is present day mean CMB temperature in the units of $\mu$ K. In the same figure, we also plot the thermal SZ \cl from hot gas in clusters of galaxies, the kinetic SZ \cl from ICM being subdominant.  
We find that SZ \cl's from CGM peak above $\ell \sim 15000$, whereas the thermal SZ from ICM peaks at $\ell \sim 3000$ and then falls at higher $\ell$-values.;
the tSZ signal from CGM dominates that from ICM over $\ell > 13000$, whereas the kSZ from galactic halos overtakes tSZ from clusters earlier at $ \ell > 10000$.


\begin{figure}
\begin{center}
\centerline{\epsfxsize=0.71\textwidth\epsfbox{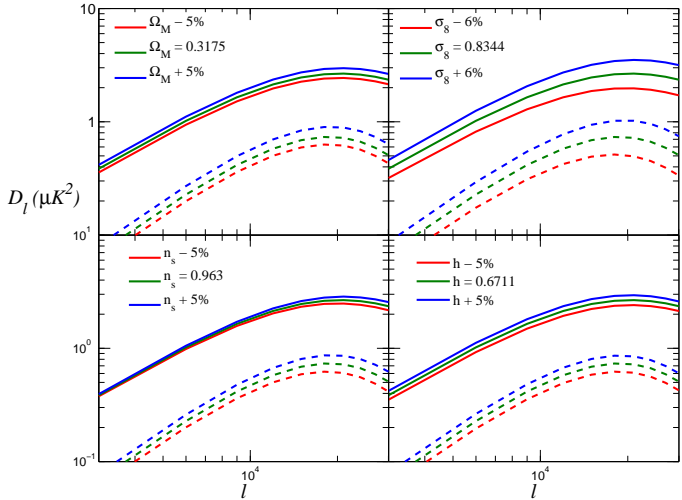}}
{\vskip-3mm}
\caption {Dependence of SZ angular power spectrum on $\sigma_8$, $\Omega_M$, $n_s$ and h. Here the dashed lines are tSZ effect and the solid lines represent
kSZ effect.}
\label{sz-params}
\end{center}
\end{figure}


We have over-plotted South Pole Telescope (SPT) and Atacama Cosmology Telescope (ACT) data, with grey and black bars, respectively and auto correlation lines from 
Figure 4 of \cite{addison12} on top of SZ \cl from galactic halos for a smaller range $3000 < \ell < 10000$ in Figure \ref{spt}.
The figure shows the contribution from thermal and kinetic SZ from galactic halos with red solid (thick) and blue solid (thick) lines.
For comparison, the thermal and kinetic SZ signals from galaxy clusters are shown as red and blue solid (thin) lines. Also, the contribution from
the sources responsible for the Cosmic Infrared Background (CIB) are shown, for both poisson (brown dashed line) and the clustered case (brown dot-dashed line). The contribution from clustering of radio sources is shown in green as a dotted line. 
The lensed primary signal is shown as a  black dashed line.
The comparison of tSZ and kSZ signals from galactic halos and galaxy clusters show that kSZ signal from galactic halos become comparable to galaxy cluster signals  at $ \ell \sim 10000$.  
This is because of the fact that kSZ is more important for lower mass halos, which correspond to smaller angles and larger $\ell$ values.

\subsection{Redshift distribution of the angular power spectrum}
The redshift distribution of $C_l$ can be determined using
\begin{equation}
 \frac{d \ln C_l}{d \ln z}=\frac{z\frac{dV}{dz}\int dM \frac{dn(M,z)}{dM} |y_l(M,z)|^2}
 {\int dz\frac{dV}{dz}\int dM \frac{dn(M,z)}{dM} |y_l(M,z)|^2}
\end{equation}
 We show the redshift distribution of $C_l$ for $\ell = 3000, 6000, 10000$ and $20000$ for tSZ and kSZ effect in Figure \ref{zdist}. For tSZ effect (shown in thin lines), for
 $\ell = 3000$, $C_l$ has a peak at $z\sim2$. This peak shifts to higher redshifts with increasing value of $\ell$. For all $\ell$ values ($\ell>3000$) there is non-negligible contribution to $C_l$ coming from $z>5$.
 
 In case of the kSZ  effect (thick lines), for $ \ell= 3000$ there is a broad peak around $z\sim1\hbox{--}2$ and the contribution to $C_l$ is significant even below $z = 1$.
 The peak shifts to higher redshifts with increasing value of $\ell$. The contribution
 from higher redshift becomes more important for larger $\ell$ values. Note that $C_l$ scales as the square of the fraction of hot gas in galactic halos,
 and the plotted values assume the fraction to be $0.11$. If the fraction is smaller, the values of $C_l$ for kSZ and tSZ are correspondingly lower. For example, if the hot halo gas constitutes only half of the missing baryons, with a fraction $\sim 0.05$ (instead of $0.1$), then SZ signal from galactic halos would dominate at 
 $ \ell \ge 30000$  (instead of $10^4$).

 \subsection{Mass distribution}
We can estimate the range of masses  which contribute most to  the thermal and kinetic SZ effects, by
computing appropriate moments of the mass function, for pressure and peculiar velocity. Figure \ref{fig:moments} shows the moment of $y-$parameters for tSZ and kSZ in the top and bottom panels, respectively,
for the mass range $10^{10} \hbox{--} 10^{13}h^{-1} M_{\odot}$, corresponding to the $l$-range $\sim 7\times10^4 \hbox{--} 7\times10^3$ for $z=1$, and 
$l$-range $\sim 1.4\times10^5 \hbox{--} 1.4 \times10^4$ for $z=4$.
The moments of tSZ ($y_{\rm tSZ} \times  {dN \over d\log M}$) show that the dominant mass range decreases with increasing redshift, from being $\sim 10^{13}$ $h^{-1}$ M$_\odot$ at $z\sim 1$, to halos of $ \sim 5\times10^{11} $ $h^{-1}$ M$_\odot$ at $z\sim 2\hbox{--}3$ to lower masses at higher redshift. From the redshift distribution
information in Figure \ref{zdist}, we can infer that galactic halos with mass $\sim 10^{12}$ $h^{-1}$ M$_\odot$ are the dominant contributors for $\ell \le 10^4$ for tSZ effect.

The moments of the kSZ signal ($y_{\rm kSZ} \times  {dN \over d\log M}$) show that low mass galactic halos are the major contributors
to the signal, and become progressively more important at increasing redshifts. Since we have constrained the mass range from a cooling time-scale argument, the moments at different redshift show that the dominant mass is $ \sim 5\times10^{11}$ $h^{-1}$ M$_{\odot}$ for $z\sim 1\hbox{--}3$. Again, from the redshift distribution information in Figure \ref{zdist}, this implies that galactic halos with $\sim 10^{12}$ $h^{-1}$ M$_\odot$ are the 
major contributors, as in the case of tSZ effect. Since significant contribution for tSZ and kSZ comes from low mass
halos, our predictions are sensitive to the assumed lower mass in which the hot halo gas can remain hot until the next merging event.

\subsection{Dependence of SZ angular power spectrum on cosmological parameters}
We also calculate the dependence of the SZ angular power spectrum on different cosmological parameters. In Figure \ref{sz-params},
we plot the dependences of tSZ and kSZ signals on $\sigma_8$, $\Omega_M$, $n_s$ and h with dashed and solid lines, respectively.
When one cosmological parameter is varied, others are kept fixex. However, when $\Omega_M$ is varied,
$\Omega_{\Lambda}$ is also changed to keep $\Omega_M+\Omega_{\Lambda}=1$.

The dependences of $C_l$ on different cosmological parameters can be fit by power-law relations near the fiducial values of the
corresponding parameters.  For example, we find that near the fiducial value of $\sigma_8$, 
  $C_l\propto {\sigma_8}^6$, which is similar to the dependence of tSZ signal from galaxy clusters \citep{komatsu02}. 
  For other parameters, we have, for tSZ, $C_l\propto {\Omega_M}^3$, $C_l\propto {n_s}^{7/2}$ and
$C_l\propto h^3$ for tSZ. The corresponding dependences for kSZ are: $C_l\propto {\sigma_8}^5$, $C_l\propto {\Omega_M}^2$, $C_l\propto n_s$ and
$C_l\propto h^2$.


\begin{figure}
\begin{center}
 \centerline{\epsfxsize=0.45\textwidth\epsfbox{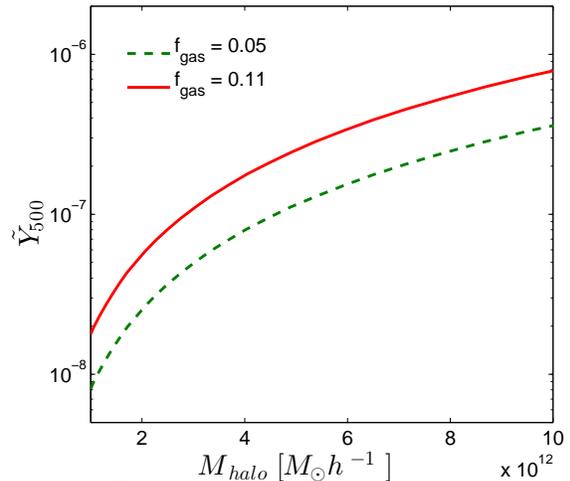}
}
{\vskip-3mm}
 \caption {$\tilde{Y}_{500}$ as a function of halo mass (red solid line) for $f_{gas}=0.11$ and $f_{gas}=0.05$ (green dashed line).}
 \label{y500}
 \end{center}
\end{figure}

\section{Detectability in future surveys and constraining gas physics}

\begin{figure}
\begin{center}
\centering
\includegraphics[width=8.0cm,angle=0 ]{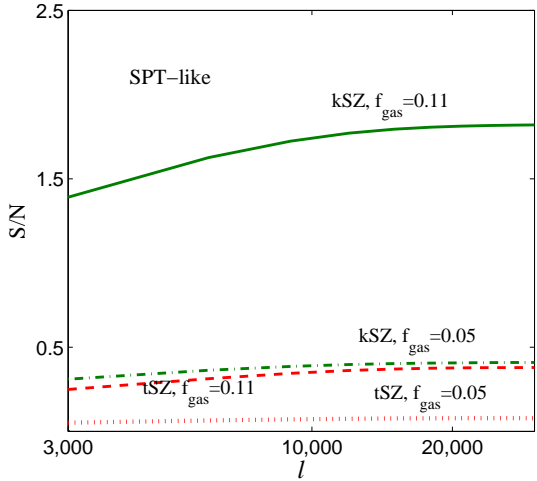}
 \caption {The cumulative signal to noise ratio (SNR) as a function of $\ell_{max}$, in case of 
 \spt survey, with $\ell_{min} = 3000$. The upper line corresponds to kSZ and the lower line to tSZ. The solid (green) and dashed (red) lines corresponds to kSZ and tSZ
 for $f_{\rm gas}=0.11$, respectively, and the dot-dashed (green) and dotted (red) lines, for $f_{\rm gas}=0.05$.}
 \label{fig:snr}
 \end{center}
\end{figure}


\begin{table}
\caption{Fiducial values and priors on the parameters}
\centering 
\begin{tabular}{l c c c c }
 \hline \hline
 Parameter & Fiducial value & \pra & \prb & \prc \\[0.5ex]
 \hline
$\sigma_8$ & 0.8344 & 0.027 & 0.027 & 0.027\\
$\Omega_M$ & 0.3175 & 0.020 &0.020 & 0.020\\
$n_s$ & 0.963 & 0.0094 & 0.0094 & 0.0094\\
$h$ & 0.6711 & 0.014 & 0.014 & 0.014\\
$f_{\rm tratio}$ & 1.0 & - & 1.0 & 1.0\\
$f_{\rm Temp}$ & 1.0 & -  & - & 0.25\\
$f_{\rm gas}$ & 0.11 & - & - & -\\
$\alpha_{\rm gas}$ & 0.0 & - & - & -\\
\hline
 \end{tabular}
\label{tab_fiducial}
\end{table}

\subsection{Integrated Comptonization parameter $\tilde{Y}_{500}$}
Next we estimate the integrated Comptonization parameter for CGM. 
The Comptonization parameter $Y_{500}$ (due to tSZ) integrated over a sphere of radius $R_{500}$ is
\begin{equation}
 Y_{500}=\frac{\sigma_T}{m_e c^2} \int_0 ^{R_{500}} \frac{PdV}{D^2_A(z)}=\frac{\sigma_T n_e k_b T_e}{m_e c^2 D^2_A(z)}\frac{4 \pi R^3_{500}}{3}
\end{equation}
where $D^2_A(z)$ is the angular diameter distance, $P=n_e k_b T_e$ is pressure of electron gas and $R_{500}$ is defined as the radius within which 
the mean mass density is $500$ times the critical density of the universe.
The second equality in the above equation is for the case of constant electron density and temperature.
The integrated Comptonization parameter scaled to z=0 is defined as
\begin{equation}
 \tilde{Y}_{500}\equiv Y_{500}E^{-2/3}(z) \Bigl( \frac{D_A(z)}{500 Mpc}\Bigr)^2 \,.
\end{equation}
Here $\tilde{Y}_{500}$ and $Y_{500}$ are expressed in square arcmin.We show in Figure \ref{y500} the values of $\tilde{Y}_{500}$ as a function of halo mass
for gas fractions $f_{0.11}$ and $f_{gas}=0.05$.
We have used the fit for concentration parameter ($c$) as a function of halo mass from \cite{duffy08}. 

From Table 1 of \cite{planck13},
the lowest stellar mass bin for which SZ signal has been detected ($\tilde{Y}_{500} \sim 10^{-6} \, {\rm arcmin}^2$) is $M_{*}\sim 4\times10^{12} M_{\odot}$. This stellar mass 
corresponds to a virial mass $\sim 4.25\times10^{12} M_{\odot} h^{-1}$. From our calculations for a galactic halo of $M_{vir}\sim 4.25\times10^{12} M_{\odot} h^{-1}$
with $f_{gas}=0.11$, the $\tilde{Y}_{500} \sim 0.2\hbox{--}0.3\times10^{-6} \, {\rm arcmin}^2$, consistent with the observed values (Table 1 of \cite{planck13}). If we use $f_{gas}=0.05$,
$\tilde{Y}_{500}$ goes down by roughly a factor of 2.



\begin{figure*}
\includegraphics[width=17.0cm,angle=0 ]{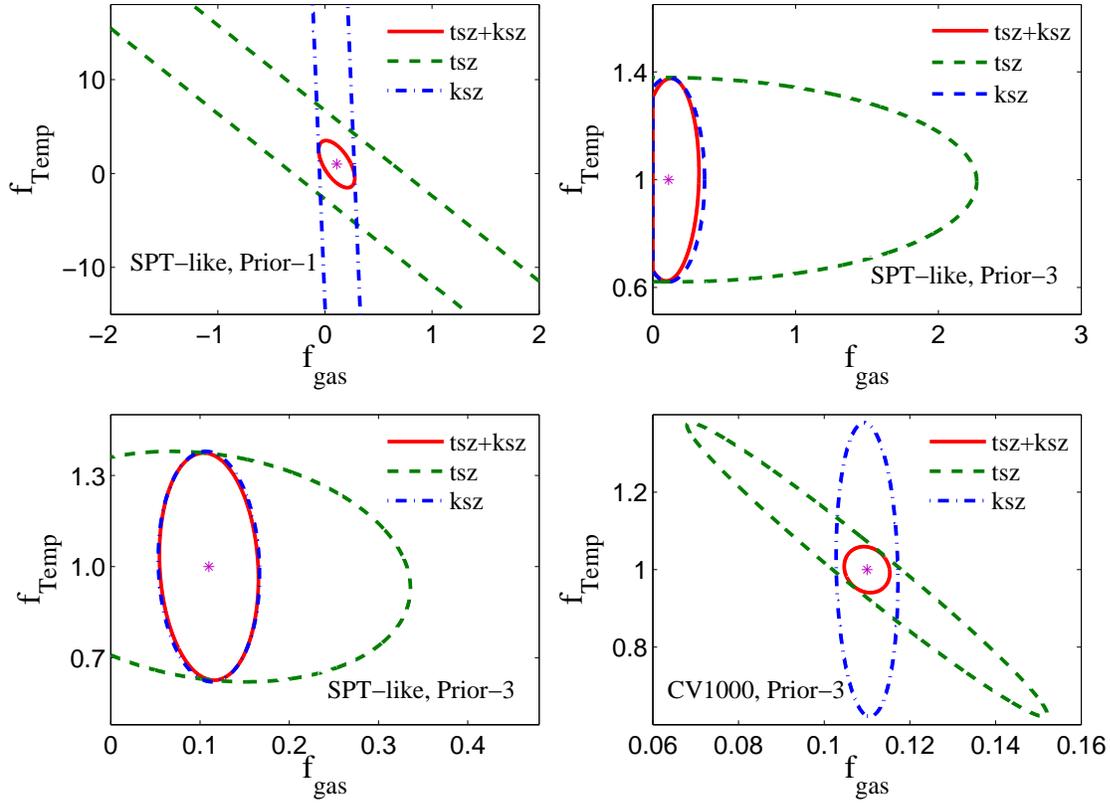}
\caption{
The figure shows the breaking of parameter degeneracy when information from tSZ \cl and kSZ \cl are taken together for different cases. The Upper Left panel shows result from the case \pra, i.e., cosmological priors only, for the \spt survey even when $\alpha$ is fixed. The Upper Right panel shows the case \prc which includes additional priors on $f_{\rm ratio}$ and $f_{\rm Temp}$ but $\alpha$, now, varied for \spt. The Lower Left panel shows  the case \prc but with $\alpha$ fixed. The Lower Right panel shows the case for \cv for, with \prc and $\alpha$ varied. In all cases, green dashed line is tSZ, blue dot-dashed is kSZ and solid red line is tSZ+kSZ.
}
\label{fig:ksztsz} 
\end{figure*}



\begin{figure*}
\includegraphics[width=8.5cm, angle=0]{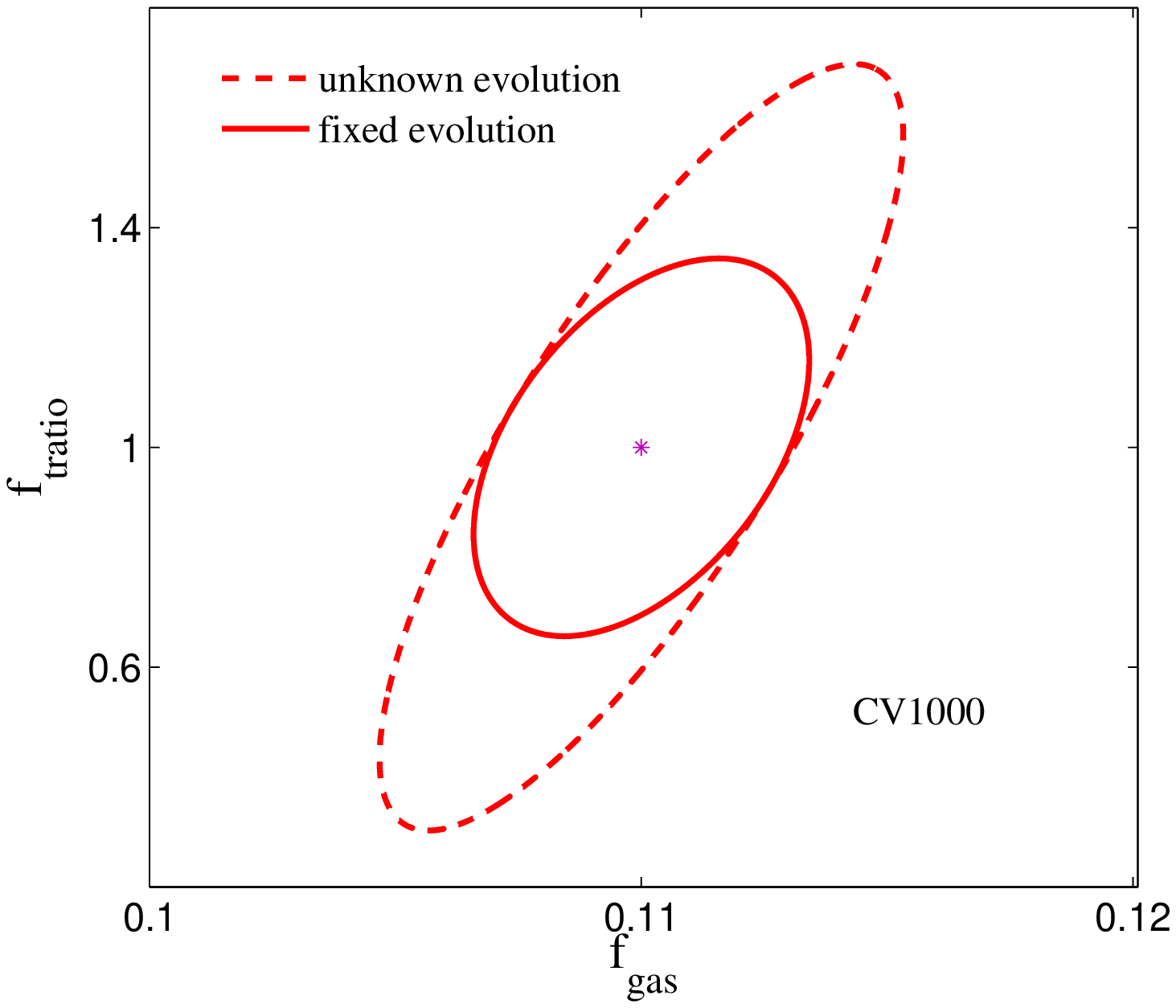}
\includegraphics[width=8.5cm, angle=0]{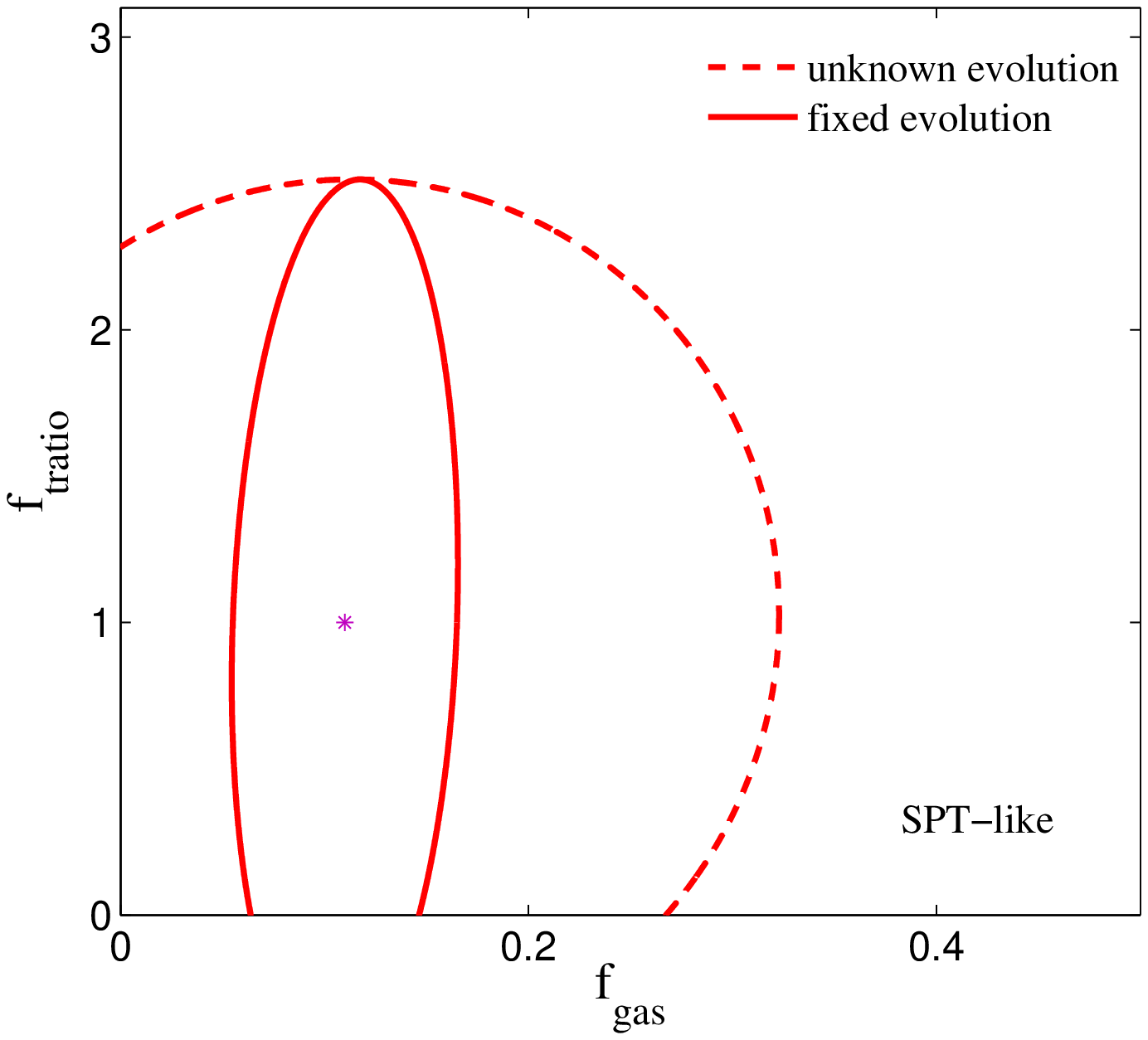}
\caption{
The figure shows the impact on parameter constraints due to any unknown evolution of the gas fraction with redshift parametrised as
$f_{\rm gas}(z)=f_{\rm gas}[E(z)]^{\alpha_{\rm gas}}$. The Left Panel is for the survey '\cv' and Right Panel is for the survey '{\sf  SPT-like}'.
In both cases, the red dashed line corresponds to the case of $\alpha_{\rm gas}$ is unknown and varied as one of the Fisher parameters, whereas the red solid line correspond to  
$\alpha_{\rm gas}$ fixed at its fiducial value.
}
\label{fig:gasevol} 
\end{figure*}


\subsection{Signal to noise ratio in future surveys}
The detectability of the CMB distortion from circum-galactic baryons can be estimated by calculating the cumulative Signal-to-Noise-Ratio (SNR) of the SZ power spectrum for a particular survey. For our purpose, we focus on two types of surveys, one which is an extension of the ongoing SPT survey to higher multipoles (although we show that the SNR from $\ell>15000$ does not add much to the cumulative SNR), and a more futuristic survey which covers 1000 square degrees of the sky (i.e, $f_{\rm sky} \sim 2\%$) and is cosmic variance error limited. These are labeled `{\sf  SPT-like}' and `\cv', respectively, for the rest of the paper:

\noindent
(1) \spt survey: In this case we use $\ell_{\rm min}$ = 3000 and $\ell_{\rm max}$ = 30000. The noise in the measurement of $C_l$'s (i.e. $\triangle C_l$) is 
taken from actual SPT data (Figure 4 of \cite{addison12}). These errors are then fitted with a power-law dependence on $\ell$ and extrapolated  till $\ell = 30000$. 

\noindent
(2) \cv survey: This survey has $2 \%$ sky coverage and the error on $C_l$'s are cosmic variance limited. Here, we have used  a smaller $\ell$-range and have taken $\ell_{\rm min}$ = 6000 and  $\ell_{\rm max}$ = 9000.

The cumulative SNR, for SZ \cl between $\ell_{\rm min}$ and $\ell_{\rm max}$, is given by
\begin{equation}
\rm{SNR_{cumu}}(\ell_{\rm min}<\ell_{\rm max}) = \left(  { \Sigma^{\ell_{\rm max}}_{ {\ell_{\rm min}}{\ell^{\prime}_{min}}}   C^X_{\ell} {(M^X_{\ell \ell{^\prime}})^{-1}} C^X_{\ell^{\prime}} } \right)^{1/2}
\label{eqn-snr}
\end{equation}
where $X$ denotes cases tSZ, kSZ or Total, i.e tSZ+kSZ and $M^X_{\ell \ell{^\prime}}$ is the corresponding covariance matrix, for any particular survey, given by
\begin{equation}
 M^X_{\ell \ell{^\prime}} \,=\, \frac{1}{4 \pi f_{\rm sky}} \left({\frac{4\pi(C^X_\ell + N_\ell)^2}{(\ell +1/2) \triangle \ell} \delta_{\ell \ell{^\prime}} + T^X_{\ell \ell{^\prime}}      }\right) \; ,
 \label{mll}
\end{equation}
where $N_\ell$ is the noise power spectrum (after foreground removal) and $T^X_{\ell \ell{^\prime}}$ is the SZ angular tri-spectrum (see, e.g., \cite{komatsu99}). Note that this formula for the covariance matrix neglects the `halo sample variance'.

The cumulative SNR provides a simple way to assess the constraining power of a given experiment irrespective of the constraints on particular parameters. We compute the cumulative SNR's for our two surveys, \spt and \cv surveys. Figure \ref{fig:snr} shows the SNR  as a function of $\ell _{\rm max}$ for the  \spt survey. Note that the covariance matrix in Equation \ref{eqn-snr}, in principle, should  include all contributions from cosmic variance (Gaussian and non-Gaussian), experimental noise after foreground removal, as well as the tri-spectrum which represents the sample variance contribution to the covariance. However, for the halo masses of interest and the $\ell$ range of the contribution of the SZ discussed in this paper, the tri-spectrum can be neglected and the covariance matrices are, effectively, diagonal. For the \cv survey, the 
diagonal covariance matrix only contains the  cosmic variance errors. 
The covariance matrix, for the \spt survey,  is taken to be  the noise (actual error bar) reported by the SPT and extrapolated to higher $\ell$'s
(as explained earlier). In general, our extrapolation of SPT errors to higher $\ell$-values are conservative in nature as  seen in Figure \ref{fig:snr} - due to the increasing observational errors for higher multipoles, the SNR for the \spt survey flattens off beyond $\ell_{max} \sim 15000$. 
It is also evident from the figure, that although it would need a stringent handle on astrophysical systematics and better modelling of SZ \cl from galaxy clusters to separate out the tSZ \cl from CGM, kSZ signal from CGM has a signal to noise ratio $\sim 2\sigma$ for the \spt survey. 
 If we take $l_{min}=10000$ for \spt survey, the signal to noise ratio goes down roughly by a factor of $2$ .
In comparison, for the more futuristc \cv survey, the tSZ and the kSZ signal can be detected with a SNR of 
$\sim 600 (950)$, at (upto) $\ell_{max} \sim 6000 (9000).$

\section{Forecasting}
\subsection{Formalism}
\label{sec-fisher}
We now employ the Fisher matrix formalism to forecast the expected constraints on the following parameters, focussing specially on the parameters related to gas physics of the circum-galactic baryons. The Fisher parameters considered are
\begin{equation}
 \{[\sigma_8, \Omega_M, n_s, h],[ f_{\rm gas},  f_{\rm tratio}, f_{\rm Temp}, \alpha_{\rm gas}] \} \;\; ,
 \label{fischer}
\end{equation}
where the first set within the parenthesis are the cosmological parameters  and the second set, which depends on baryonic physics, are the {\it astrophysical} parameters.

To construct the Fisher Matrices for the two surveys, we compute the derivatives of the tSZ, kSZ and, hence, total SZ \cl with respect to each parameter around the fiducial values listed in Table 1.
Here $f_{\rm gas}$ is the redshift independent fraction of halo mass in gaseous form  and $\alpha_{\rm gas}$ captures any possible evolution of the gas defined through
$f_{\rm gas}(z)=f_{\rm gas}[E(z)]^{\alpha_{\rm gas}}$. Our fiducial model assumes no evolution of the gas fraction; see details in section \ref{sec:ksztsz}. The other two parameters that encapsulate the uncertainty in our knowledge of hot gas in galactic halos are  
$f_{\rm tratio} = \frac{t_{\rm cool}}{t_{\rm dest}}$, i.e.,  the ratio of cooling time to destruction time for galactic halos, $f_{\rm Temp} = \frac{T}{T_{\rm vir}}$, i.e.,  the ratio of the temperature of the gas to the virial temperature of gas in a halo.  

For a given fiducial model, the Fisher matrix is written as
\begin{equation}
 F_{ij} \,=\,\frac{\partial C^X_\ell}{\partial p_i} (M^X_{\ell \ell{^\prime}})^{-1} \frac{\partial C^X_{\ell{^\prime}}}{\partial p_j}
\end{equation}
where $M_{ll'}$ is given by equation \ref{mll} in case of \cv survey and for \spt survey we have $M_{ll} = (\Delta C^{SPT}_\ell)^2$.
Here $\Delta C^{SPT}_\ell$'s are the error on $C_\ell$'s from SPT data. 
The fiducial values and the priors used are listed in Table \ref{tab_fiducial}. Note that in all our calculations, cosmological priors are always applied. Priors related to gas/halo physics are additionally applied, on a case by case basis.  
For the rest of the paper, we denote the different priors uses as follows:\\
\pra:  Priors on cosmological parameters only.\\
\prb:  Priors on cosmological parameters + 100\% prior on $f_{\rm tratio}$.\\
\prc:  Priors on cosmological parameters + 100\% prior on $f_{\rm tratio}$ + 25\% prior on $f_{\rm Temp}$.\\

In \prc and \prb, we have assumed a 100\% prior on $f_{\rm tratio}$, reflecting the maximum uncertainty in this parameter. For $f_{\rm Temp}$, we have assumed a smaller
uncertainty, since our constraint that cooling time is longer than the destruction time ensures that the gas temperature to be close to the
virial temperature.

Additionally, for each case considered, we look at constraints for  all the 8 parameters listed above (equation \ref{fischer}) and in the
second case, we repeat the same procedure but with only 7 parameters, assuming that the baryonic content of galaxies is independent of redshift (i.e. $\alpha_{\rm gas}=0$). The introduction of varying gas fraction in halos changes the shape of \cl (see, for example, in \cite{majumdar01})which results in different sensitivity to the Fisher parameters; it also introduces an extra nuisance parameter to be marginalised over. The results of the first analysis (with $\alpha_{\rm gas}$ varying) are shown in Table 3 and the second case (with $\alpha_{\rm gas}$ fixed)
in Table 2.

\subsection{Results}
We are in an era in cosmology where major surveys like {\it Planck} have already provided tight constraints on the parameters of the standard cosmological model. In the future, two of the major goals are to go beyond the standard model of cosmology and to constrain parameters related to baryonic/gas physics associated with non-linear structures. One of the puzzles related to baryonic matter is the issue of 'missing baryons', i.e the fact that after accounting for the gas locked up in structures (like galaxies and galaxy clusters) and the diffuse intergalactic medium, one still falls short of the cosmological mean baryon fraction $\Omega_B$. While recently, much of this missing material may  have been accounted by the intra-cluster medium, a deficit of the order of at least $\sim 10$'s \% is still found. 

With the growing observational evidence for CGM, it would be interesting to determine if its inclusion  in the baryon census can fill the deficit. To go forward, one needs to go beyond the discovery of the CGM in nearby 
isolated halos (other than the Milky Way) or beyond what one can measure by doing a stacking analysis of gas in a sample of halos. This is possible by probing the locked gas in and around  a cosmological distribution of galaxy halos through its signature on the CMB as shown in this paper.  A constraint on the mean gas fraction, $f_{\rm gas}$, included in our calculations, provides one of the  best ways to estimate the amount of circum-galactic baryons in a statistical sense. In the rest of the section, we focus on the constraints on $f_{\rm gas}$, for a variety of survey scenarios.

\begin{figure*}
\includegraphics[width=17.0cm,angle=0 ]{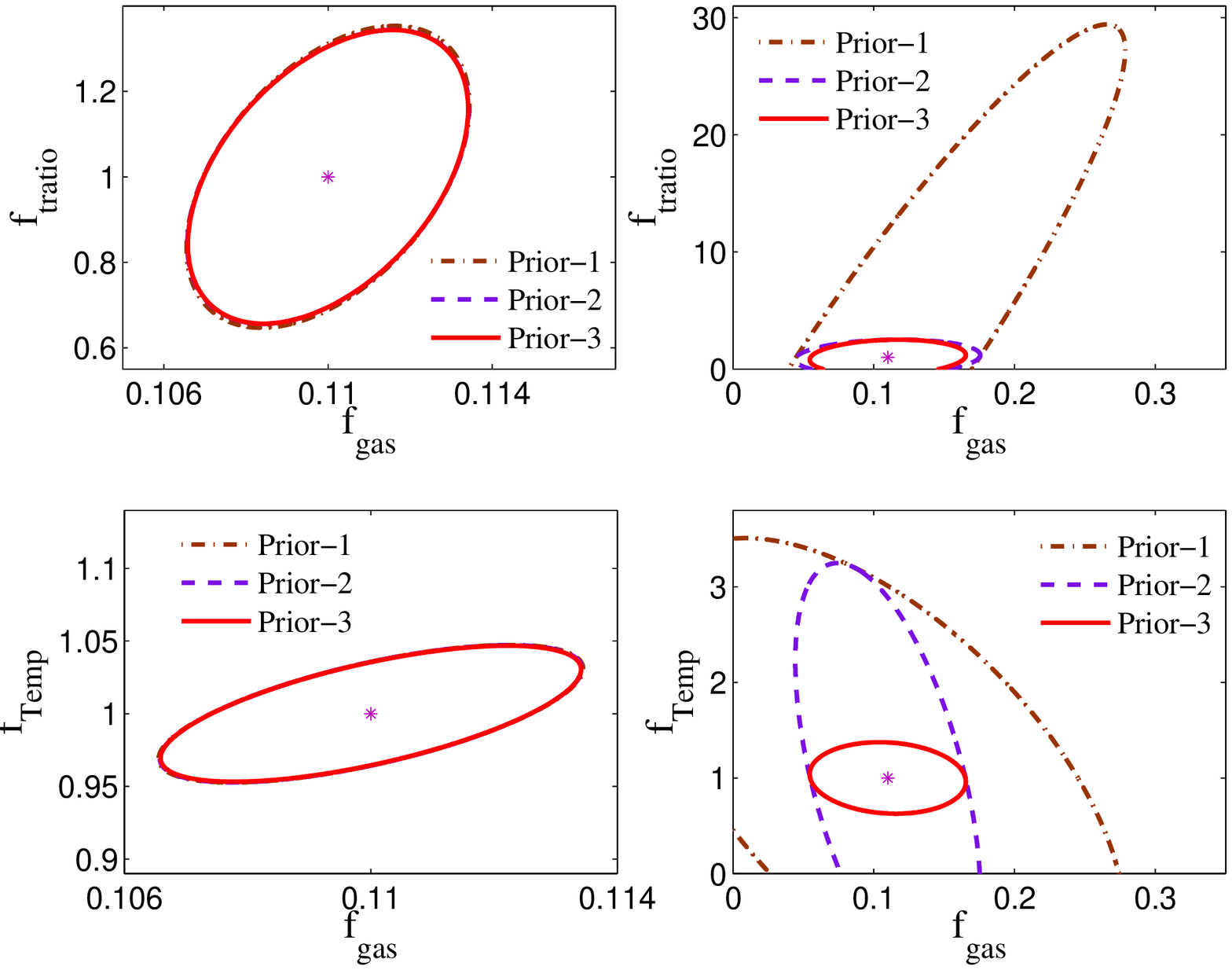}
\caption {$1$-$\sigma$ contours for gas physics parameters $f_{\rm gas}, f_{\rm tratio}, f_{\rm Temp}$ when $\alpha_{\rm gas}$ is fixed. The left panel is for \cv survey and the right panel is for {\sf  SPT-like} survey. In all cases red solid line is Prior-3, purple dashed line is Prior-2 and brown dot-dashed line is Prior-1.
}
 \label{fig:7params}
\end{figure*}

\begin{figure*}
\includegraphics[width=17.0cm,angle=0 ]{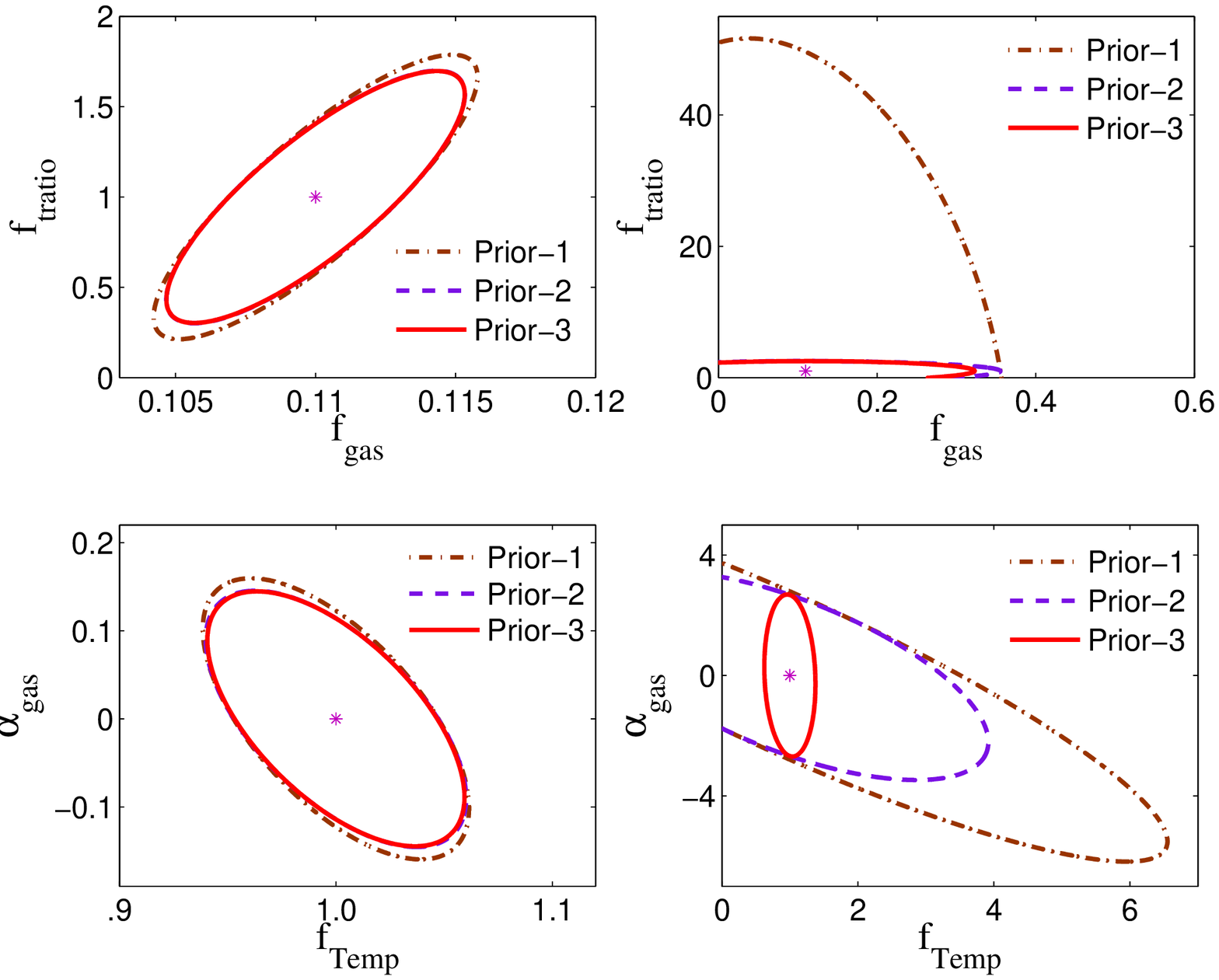}
\caption {$1$-$\sigma$ contours for gas physics parameters $f_{\rm gas}, f_{\rm tratio}, f_{\rm Temp}, \alpha_{\rm gas}$. The left panel is for \cv survey and the right panel is for {\sf  SPT-like} survey.  In all cases red solid line is Prior-3, purple dashed line is Prior-2 and brown dot-dashed line is Prior-1.
}
 \label{fig:8params}
\end{figure*}

The constraints on the amount of baryons locked up as CGM, as well on other Fisher parameters, are shown in Tables \ref{tab:7params} \& \ref{tab:8params}. The $1\sigma$ ellipses for joint constraints of \fg with non-cosmological parameters, for the two surveys considered and different prior choices, are shown in Figures \ref{fig:7params} $-$ \ref{fig:8params}. 


\begin{table*}
\caption{Error on parameters for different surveys and Prior cases with fixed $\alpha_{\rm gas}$}
\centering 
\begin{tabular}{l l l l l l l}
\hline
 Parameters & \cv, P1 & \cv, P2 & \cv, P3 & \spt, P1 & \spt, P2 & \spt, P3 \\
 \hline
 $\Delta \sigma_8$ & 0.0166 & 0.0163 & 0.0162 & 0.0270 & 0.0270 & 0.0270 \\[-0.5ex]
 $\Delta \Omega_M$ & 0.0163 & 0.0161 & 0.0161 & 0.020 & 0.020 & 0.020 \\[-0.5ex]
 $\Delta n_s$ & 0.0093 & 0.0093 & 0.0093 & 0.0094 & 0.0094 & 0.0094 \\[-0.5ex]
 $\Delta h$ & 0.0139 & 0.0139 & 0.0139 & 0.0140 & 0.0140 & 0.0140 \\[-0.5ex]
 $\Delta f_{\rm tratio}$ & 0.2329 & 0.2268 & 0.2266 & 18.7380 & 0.9986 & 0.9982 \\[-0.5ex]
 $\Delta f_{\rm Temp}$ & 0.0312 & 0.0311 & 0.0309 & 1.6547 & 1.4826 & 0.2465 \\[-0.5ex]
 $\Delta f_{\rm gas}$ & 0.0023 & 0.0023 & 0.0023 & 0.1119 & 0.0433 & 0.0366 \\[-0.5ex]
 \hline
 \end{tabular}
 \label{tab:7params}
\end{table*}



\begin{table*}
\caption{Error on parameters for different surveys and Prior cases}
\centering 
\begin{tabular}{l l l l l l l}
\hline
 Parameters & \cv, P1 & \cv, P2 & \cv, P3 & \spt, P1 & \spt, P2 & \spt, P3 \\
 \hline
 $\Delta \sigma_8$ & 0.0270 & 0.0263 & 0.0261 & 0.0270 & 0.0270 & 0.0270 \\[-0.5ex]
 $\Delta \Omega_M$ & 0.020 & 0.0187 & 0.0187 & 0.020 & 0.020 & 0.020 \\[-0.5ex]
 $\Delta n_s$ & 0.0094 & 0.0094 & 0.0094 & 0.0094 &0.0094 & 0.0094 \\[-0.5ex]
 $\Delta h$ & 0.0140 &  0.0140 & 0.0140 & 0.0140 & 0.0140 & 0.0140 \\[-0.5ex]
 $\Delta f_{\rm tratio}$ & 0.5192 & 0.4608 & 0.4598 & 33.393 & 0.9995 & 0.9984 \\[-0.5ex]
 $\Delta f_{\rm Temp}$ & 0.0405 & 0.0396 & 0.0392 & 3.6606 & 1.9240 & 0.2479 \\[-0.5ex]
 $\Delta f_{\rm gas}$ & 0.0038 & 0.0035 & 0.0035 & 0.1687 & 0.1619 & 0.1404 \\[-0.5ex]
 $\Delta \alpha_{\rm gas}$ & 0.1052 & 0.0958 & 0.0954 & 4.0753 & 2.2890 & 1.7734 \\[-0.5ex]
 \hline
 \end{tabular}
 \label{tab:8params}
\end{table*}



\subsubsection{Constraints on CGM using using kSZ + tSZ}
\label{sec:ksztsz}
Strong degeneracies between the astrophysical parameters prevent us from getting any useful constraints on the CGM, using {\it only} cosmological priors i.e \pra, when one uses either of the tSZ or the kSZ \cl alone. However, once both the tSZ and the kSZ signals are added, the strong degeneracies are broken. This is seen clearly in the upper left panel of Figure \ref{fig:ksztsz}, which shows the joint constraint for the \spt survey. The fact that two cigar-like degeneracies, from two datasets, differing in their degeneracy directions eventually  leads to very strong constraints in parameter space when taken together, is well known (see, for example, \cite{khedekar10}) and the same idea is at work here. Thus, although there is practically no constraint on \fg from using tSZ or kSZ \cl from CGM individually, adding them together results in a weak constraint of 
$\Delta f_{\rm gas} \approx 0.11$ which is the same as the fiducial value of \fg.  One of reason for this weak constraint is the additional degeneracy of \fg with $\alpha$.

 This degeneracy of \fg with $\alpha$ is broken either (i) when one evokes no evolution in the Fisher analysis or (ii) when additional astrophysical priors are imposed. This is shown in the upper right and lower left panels of Figure \ref{fig:ksztsz}. In both cases, the addition of astrophysical priors, for example \prc, can already break the strong cigar like degeneracies leaving both kSZ and tSZ signal power to constrain \fg. The difference between these two panels is that $\alpha$ is not fixed (i.e we marginalise over unknown evolution) for the upper right panel leading to slightly weaker constraints (for tSZ+kSZ) than the lower left panel where $\alpha$ is held constant. The higher SNR of kSZ w.r.t tSZ (as seen in Figure \ref{fig:snr}) gives the kSZ \cl a stronger constraining power on \fg than tSZ and the addition of tSZ \cl  makes only modest improvement on the constraint on CGM achieved by using kSZ \cl only. 

The lower right panel of Figure \ref{fig:ksztsz} shows that constraints from the more futuristic cosmic variance limited survey \cv in the presence of \prc but including an unknown gas fraction. In this case, due to its better sensitivity, tSZ is capable of constraining \fg (compare green dashed ellipses in the two right panels, upper and lower) and finally comes up with stronger joint constraint than \spt (compare the red solid ellipse in lower left and and lower right). In the rest of this section, we focus mainly on constraints coming from kSZ+tSZ \cl,  keeping  in mind that all the constraints will only be slightly degraded if  only kSZ \cl are used instead. Note that this is applicable as long as the astrophysical priors are added. 

As evident above, one of the major uncertainties in our knowledge of the gas content of halos at all scales is our lack of understanding of any redshift evolution of the gas. In using large-scale structure data to constrain cosmology, for example, an unknown redshift evolution can seriously degrade cosmological constraints (as an example, see \cite{MM03}) and one needs to invoke novel ideas to improve constraints \citep{MM04, khedekar13}. Whereas for galaxy clusters, in which case \fg has been measured at higher redshift, and one finds evolution in gas content,  no such evolution has been measured for galactic halos considered in this work. It is 
however possible that feedback processes in galaxies, and cosmological infall of matter may introduce an evolution of \fg with redshift. 
 In order to incorporate the impact of gas evolution on our constraints, we have considered the 
possibility that  \fg to scales with the expansion history $E(z)$ with a power-law index $\alpha$,  with the fiducial value of $\alpha$ set to 0.  

The constraints on all the parameters used in the Fisher analysis for the cases where we assume the gas fraction to remain constant are given in Table \ref{tab:7params}.    As mentioned before, in the absence of any astrophysical priors, there is no interesting constraints on \fg (as well as \ftemp or \fratio) for \spt survey. However, for the \cv survey the amount of gas locked as CGM can be constrained very tightly to better than  2\%; similarly, with cosmological priors only \cv can constraint departure from the virial temperature to  3.1\% and \fratio to $ \sim 23$\%.  The addition of astrophysical priors, either \prb or \prc does not improve the constraints for \cv any further, since the constraints with \pra are much tighter than the priors imposed. However, astrophysical priors considerably improve the constraints for the  \spt survey especially for {\it \fg which is constrained to  39\% when \prb is used and is further constrained to better than  33\% accuracy with \prc}. This means that for both \prb and \prc, \
fg=0 can be excluded by at least $3\sigma$ with the \spt survey.

The corresponding constraint ellipses showing the $1\sigma$ allowed region between \fg and either \ftemp or \fratio are shown in Figure \ref{fig:7params}. The left panels show the degeneracy ellipses for \cv whereas the right panels show the same for \spt. Notice, from the upper panels, that  \fratio has a positive correlation with \fg. This can be understood by noting that any increase in \fg increases \cl whereas it can be offset by an increase \fratio which pushes up the lowe limit of halo mass (see Figure \ref{cool}) and hence decreases the number density of halos thus lowering the \cl. The anti-correlation of \fg with \ftemp, seen in the lower right panel, is a consequence of the anti-correlation of $n_e$ and $T_v$ (in Equation \ref{eqn-ytheta}) in the tSZ relation which modulates the overall degeneracy direction of tSZ+kSZ. Note that for the  \cv survey, the $1\sigma$ ellipses are almost degenerate whereas priors shape the relative areas of the ellipses for the \spt survey.

A fixed non-evolving \fg, although desirable, is rather naive. Given our lack of understanding of the the energetics affecting the CGM over cosmic time scales, it is prudent to marginalise over any unknown evolution of \fg parametrised, here, by $\alpha$. The resultant constraints are given in Table \ref{tab:8params}.  The presence over one extra unknown gas evolution parameter to marginalise over dilutes the constraints on \fg for the both the surveys. 
For the \cv survey, the constraints are still strong and hovers around  3\%  for all the three prior choices. Moreover, \ftemp and \fratio can be still be constrained to $ \sim 4\%$  and $ \sim 46\%$  by the futuristic survey.
 Without any external prior on $\alpha$, all parameters poorly constrained by the \spt survey. With \cv survey, one can get a much stringent constraint on any possible evolution of the CGM with $\Delta \alpha \sim 0.1$

\subsubsection{Constraints on Cosmology}
The  parameters of the standard cosmological model are already tightly constrained by {\it Planck}. These are the constrains that are used as \pra in this paper.  With the SNR possible in a \spt survey, it is not possible to tighten the cosmological constraints further irrespective of whether we know $\alpha$ or it is marginalized over. However, with the larger sensitivity of \cv survey, it is possible to further improve cosmological parameters, albeit with $\alpha$ fixed. A quick look at Table \ref{tab:7params} shows that it is possible to shrink the $1\sigma$ error on $\sigma_8$  by almost a factor of 2 and that on $\Omega_M$ by $ \sim 20 \%$.

\subsubsection{Constraints on the density profile of CGM}
We have so far assumed the density profile of CGM to be uniform, which was argues on basis of current observations \citep{putman12, gatto13}. However, it is perhaps more realistic to assume that the density profile to decrease at large galacto-centric distances. One can ask if it would be possible to determine the pressure profile of the halo gas from SZ observations in the near future. In order to investigate this, we parameterise the density profile by $\gamma_{gas}$ such that $\rho_{gas}(r) \propto {(1+(\frac{r}{R_s})^{\gamma_{gas}})}^{-1}$,
where $R_s$ is the scale radius defined as $R_s \equiv R_{vir}/c(M,Z)$ and $c(M,Z)$ is the concentration parameter. This density
profile gives uniform density at $r<<R_s$ and $\rho_{gas}(r) \propto {r^{-\gamma_{gas}}}$ at $r>>R_s$. We include $\gamma_{gas}$ in Fisher 
matrix analysis with fiducial value $\gamma_{gas}=0$. For \cv survey with a fixed $\alpha_{gas}$ and \prc, the constrain on 
density profile of CGM is $\gamma_{gas}<1.5$ whereas the constrain degrades to $\gamma_{gas}<3.15$ in the presence of an unknown
redshift evolution of gas fraction. $\gamma_{gas}$ poorly constrained by \spt survey.

\section{SZ effect from warm CGM}
The observations of \cite{tumlinson11} have shown the existence of OVI absorbing clouds, at $10^{5.5}$ K, with hydrogen column density 
$N_{\rm H} \sim 10^{19\hbox{--}20}$ cm$^{-2}$. The integrated pressure from this component in the galactic halo is estimated as $\langle p \rangle \sim N_{\rm H} k T$. This implies a thermal SZ $y-$ distortion of order $y_{OVI} \sim  N_{\rm H} kT \sigma_T/(m_e c^2) \sim 3.6 \times 10^{-9} \, N_{\rm H, 20}$, where $N_{\rm H}=10^{20} N_{\rm H,20}$ cm$^{-2}$.

There is also a cooler component of CGM, at $\sim 10^4$ K, which is likely to be in pressure equilibrium with the warm CGM. 
The COS-Halos survey have shown that a substantial fraction of the CGM can be in the form of cold ($\sim 10^4$ K). Together 
with the warm OVI absorbing component, this phase can constitute more than half the missing baryons \citep{werk14}.
Simulations of the interactions of galactic outflows with halo gas in Milky Way type galaxies also show that the interaction zone suffers from various instabilities, and forms clumps of gas at $10^4$ K (\cite{marinacci10, sharma14}). These are possible candidates of clouds observed with NaI or MgII absorptions in galactic halos. Cross-correlating MgII absorbers with SDSS, WISE and GALEX surveys,
\cite{lan14} have concluded that some of the cold MgII absorbers are likely associated with outflowing material.
However, for similar column density of these clouds, the SZ signal would be less than that of the warm components by $10^{-1.5}$ because of the temperature factor.

We can calculate the integrated $y$-distortion due to the CGM in intervening galaxies, by estimating the average number of galaxies
in the appropriate mass range ($10^{12\hbox{--}13} $ M$_\odot$) in a typical line of sight, using Monte-Carlo simulations.  Dividing
a randomly chosen line-of-sight, we divide it in redshift bins up to $z=8$, and 
each redshift bin is then populated with halos using the Sheth-Tormen mass function, in the above mentioned mass range. We estimate
the average number to be $\sim 20$ after averaging over $50$ realisations. This implies an integrated $y$-parameter of order $7.5 \times 10^{-8} \, N_{\rm H,20}$.  This can be detected with upcoming experiments such as Primordial Inflation Explorer ({\it PIXIE}) even with $N_{\rm H}=10^{19}$ cm$^{-2}$, since it aims to 
detect spectral distortion down to $y \ge 2 \times 10^{-9}$ \cite{pixie11}.

The kinetic SZ signal from the warm gas in galactic halos can be estimated from eqn \ref{eq:ksztotsz}, writing $v_{\rm local}$ as the local (line of sight) velocity dispersion. Recent studies indicate that CGM gas is likely turbulent, probably driven by the gas outflows \citep{evoli11}. If we consider transonic turbulence for this gas, then $v_{\rm los}/c \sim \sqrt{kT/m_p c^2} $. Then we have,
\begin{equation}
{\Delta T_{\rm kin} \over \Delta T_{\rm th}} \approx {1 \over 2} {m_e \over m_p} \sqrt{{m_p c^2 \over kT_e}} = {m_e \over 2 m_p} {c \over v_{\rm loc}} \,.
\end{equation}
For $v_{\rm loc}\sim 100$ km s$^{-1}$ (corresponding to gas with temperature $\sim 10^6$ K), the kSZ signal from turbulent gas is, therefore, comparable to the tSZ signal.

\section{Conclusions}
We have calculated the SZ distortion from galactic halos containing warm and hot circumgalactic gas. For the hot halo gas,
we have calculated the angular power spectrum of the distortion caused by halos in which the gas cooling time is longer than the halo
destruction time-scale (galactic halos in the mass range of $ 5 \times 10^{11} \hbox {--}10^{13} h^{-1}$ M$_\odot$. The SZ distortion signal
is shown to be significant at small angular scales ($\ell \sim 10^4$), and larger than the signal from galaxy clusters. The kinetic
SZ signal is found to dominate over the thermal SZ signal for galactic halos, and also over the  thermal SZ signal from galaxy clusters
for $ \ell > 10000$.  We also show that the estimated Comptonization parameter $\tilde{Y}_{500}$ for most massive galaxies (halo mass $\ge 10^{12.5}$ M$_\odot$) is consistent with the marginal detection by {\it Planck}.
The integrated Compton distortion from the warm CGM is estimated to be $y\sim 10^{-8}$, within the capabilities of future experiments. 

Finally, we have investigated the detectability of the SZ signal 
for two surveys, one which is a simple extension of the SPT survey that we call \spt and a more futuristic cosmic variance limited survey termed \cv. We find that for the  \spt survey,  
kSZ from CGM has a SNR of $\sim$ 2$\sigma$ and at much higher SNR for the \cv survey. We do a Fisher analysis to assess the capability of these surveys to constrain the amount of CGM. Marginalizing 
over cosmological parameters, with {\it Planck} priors, and astrophysical parameters affecting the SZ \cl from CGM, we find that in the absence of any redshift evolution of the gas fraction, the \spt survey can constrain 
\fg to $ \sim 33$\%, and the \cv survey, to $ \sim 2$\%. Solving simultaneously  for an unknown evolution of the gas fraction, the resultant constraints for 
\cv becomes 3\%  and it is poorly constrained by \spt survey. We also find that a 
survey like \cv can improve cosmological errors on $\sigma_8$ obtained by {\it Planck} by a factor of 2, if one has knowledge of the gas evolution. The Fisher 
analysis tells us that if indeed $\sim 10$\% of the halo mass is in the circumgalactic medium, then this fraction can be measured with sufficient precision and can be included in the baryonic census of our Universe.

\bigskip
PS and BN would like to thank Jasjeet Singh Bagla and Suman Bhattacharya for helpful discussions. SM acknowledges the hospitality of Institute for Astronomy at ETH-Zurich where the project was completed during the the author's Sabbatical. 

\footnotesize{

}

\begin{thebibliography}{}

\bibitem[\protect\citeauthoryear{Addison \etal}{2012}]{addison12}
Addison, G. E., Dunkley, J., Spergel, D. N. 2012, MNRAS, 427, 1741

\bibitem[\protect\citeauthoryear{Anderson \& Bregman}{2010}]{anderson10}
Anderson, M. E., Bregman, J. N. 2010, ApJ, 714, 320


\bibitem[\protect\citeauthoryear{Anderson \& Bregman}{2011}]{anderson11}
Anderson, M. E., Bregman, J. N. 2011, ApJ, 737, 22

\bibitem[\protect\citeauthoryear{Anderson \etal}{2013}]{anderson13}
Anderson, M. E., Bregman, J. N., Dai, X. 2013, ApJ, 762, 106

\bibitem[\protect\citeauthoryear{Anderson \etal}{2014}]{anderson14}
Anderson, M. E., Gaspari, M., White, S. D. M., Wang, M., Dai, W. 2014, arXiv:1409.6965v1

\bibitem[\protect\citeauthoryear{Benson \etal}{2000}]{benson00}
Benson, A. J., Bower, R. G., Frenk, C. S., White, S. D. M. 2000, 314, 557

\bibitem[\protect\citeauthoryear{Bhattacharya \& Kosowsky}{2008}]{bhattacharya08}
Bhattacharya, S., Kosowsky, A. 2008, PhRvD, 3004B

\bibitem[\protect\citeauthoryear{Birnboim \& Dekel}{2003}]{birnboim03}
Birnboim, Y., Dekel, A. 2003, MNRAS, 345, 349

\bibitem[\protect\citeauthoryear{Bogd\'an \etal}{2011}]{bogdan11}
Bogd\'an, \'A, Gilfanov, M. 2011, MNRAS, 418, 1901

\bibitem[\protect\citeauthoryear{Bogd\'an \etal}{2013a}]{bogdan13a}
Bogd\'an, \'A, Forman, W. R., Vogelsberger, M. \etal 2013, ApJ, 772, 97

\bibitem[\protect\citeauthoryear{Bogd\'an \etal}{2013b}]{bogdan13b}
Bogd\'an, \'A, Forman, W. R., Kraft, R. P., Jones, C. 2013, ApJ, 772, 98

\bibitem[\protect\citeauthoryear{Crain \etal}{2010}]{crain10}
Crain, R. A., McCarthy, I. G., Frenk, C. S., Theuns, T., \& Schaye, J. 2010, MNRAS, 407, 1403


\bibitem[\protect\citeauthoryear{Dai \etal}{2012}]{dai12}
Dai, X., Anderson, M. E., Bregman, J. N., Miller, J. M. 2012, ApJ, 755, 107

\bibitem[\protect\citeauthoryear{Duffy \etal}{2008}]{duffy08}
Duffy, A. R., Battye, R. A., Davies, R. D., Moss, A., Wilkinson, P. N. 2008, MNRAS, 383, 150

\bibitem[\protect\citeauthoryear{Dutton \etal}{2010}]{dutton10}
Dutton, A. A., Conroy, C., vanden Bosch, F. C., Prada, F., More, S. 2010, MNRAS, 407, 2D

\bibitem[\protect\citeauthoryear{Efstathiou \& Migliaccio}{2012}]{efstathiou12}
Efstathiou, G., Migliaccio, M. 2012, MNRAS, 423, 2492

\bibitem[\protect\citeauthoryear{Evoli \& Ferrara}{2011}]{evoli11}
Evoli, C., Ferrara, A. 2011, MNRAS, 413, 2721


\bibitem[\protect\citeauthoryear{Fang \etal}{2013}]{fang13}
Fang, T., Bullock, J., Boylan-Kolchin, M. 2013, ApJ, 762, 20

\bibitem[\protect\citeauthoryear{Forman \etal}{1985}]{forman85}
Forman, W., Jones, C., Tucker, W. 1985, ApJ, 293, 102

\bibitem[\protect\citeauthoryear{Fukugita \etal}{1998}]{fukugita98}
Fukugita, M., Hogan, C. J., Peebles, P. J. E., 1998, ApJ, 503, 518

\bibitem[\protect\citeauthoryear{Gatto \etal}{2013}]{gatto13} Gatto, A., Fraternali, 
F., Read, J.~I., et al.\ 2013, MNRAS, 433, 2749 



\bibitem[\protect\citeauthoryear{Gradshteyn \& Ryzhik}{1990}]{gradshteyn80}
Gradshteyn, I. S., Ryzhik, I. M. 1980, Tables of Integrals, Series and Products (New York: Academic Press)

\bibitem[\protect\citeauthoryear{Grcevich \& Putman}{2009}]{putman09}
Grcevich, J., Putman, M. E. 2009, ApJ, 696, 385

\bibitem[\protect\citeauthoryear{Hamana \etal}{2003}]{hamana03}
Hamana, T., Kayo, I., Yoshida, N., Suto, Y., Jing, Y. P. 2003, MNRAS, 343, 1312H

\bibitem[\protect\citeauthoryear{Jing}{1999}]{jing99}
Jing, Y. P. 1999, ApJ, 515, L45

\bibitem[\protect\citeauthoryear{Khedekar, Majumdar \& Das}{2010}]{khedekar10}
Khedekar, S., Majumdar, S., \& Das, S., 2010, PRD, 82, 041301

\bibitem[\protect\citeauthoryear{Khedekar \& Majumdar}{2013}]{khedekar13}
Khedekar, S., \&  Majumdar, S., S., 2013, JCAP, 2, 30


\bibitem[\protect\citeauthoryear{Kogut \etal}{2011}]{pixie11}
Kogut, A, Fixsen, D. J., Chuss, D. T. \etal 2011, JCAP, 7, 25

\bibitem[\protect\citeauthoryear{Komatsu \& Kitayama}{1999}]{komatsu99}
Komatsu, E., Kitayama, T. 1999, ApJ, 526L, 1K

\bibitem[\protect\citeauthoryear{Komatsu \& Seljak}{2002}]{komatsu02}
Komatsu, E., Seljak, U. 2002, MNRAS, 336, 1256

\bibitem[\protect\citeauthoryear{Lacey \& Cole}{1993}]{lacey93}
Lacey, C., Cole, S. 1993, MNRAS, 262, 627

\bibitem[\protect\citeauthoryear{Lacey \& Cole}{1994}]{lacey94}
Lacey, C., Cole, S. 1994, MNRAS, 271, 676

\bibitem[\protect\citeauthoryear{Lan \etal}{2014}]{lan14}
Lan, T.-W., M\'enard, B., Zhu, G. 2014, arXiv:1404.5301

 \bibitem[\protect\citeauthoryear{Leauthaud \etal}{2010}]{leau10}
Leauthaud, A., Tinker, J., Bundy, K., et al. 2012, ApJ, 744,159

\bibitem[\protect\citeauthoryear{Maccio \etal}{2007}]{maccio07}
Maccio A. V., Dutton A. A., Bosch F. C. van den, Moore B., Potter D., Stadel J., 2007, MNRAS, 378, 55

\bibitem[\protect\citeauthoryear{Majumdar}{2001}]{majumdar01}
Majumdar, S., 2001, ApJ, 555, L7

\bibitem[\protect\citeauthoryear{Majumdar \& Mohr}{2003}]{MM03}
Majumdar, S., \& Mohr, J.\,J., 2003, ApJ, 585, 603

\bibitem[\protect\citeauthoryear{Majumdar \& Mohr}{2004}]{MM04}
Majumdar, S., \& Mohr, J.\,J., 2004, ApJ, 613, 41

\bibitem[\protect\citeauthoryear{Marinacci \etal}{2010}]{marinacci10}
Marinacci, F., Binney, J., Fraternali, F., Nipoti, C., Ciotti, L., Londrillo, P. 2010, MNRAS, 404, 1464

\bibitem[\protect\citeauthoryear{Maller \& Bullock}{2004}]{maller04}
Maller, A. H., Bullock, J. S., 2004, MNRAS, 355, 694

\bibitem[\protect\citeauthoryear{Mitra \etal}{2011}]{mitra11}
Mitra, S., Kulkarni, G., Bagla, J. S., Yadav, J. K. 2011, BASI, 39, 563 

\bibitem[\protect\citeauthoryear{Mo \etal}{1998}]{mo98}
Mo, H. J., Mao, S., White, S. D. M. 1998, MNRAS, 295, 319

\bibitem[\protect\citeauthoryear{Moster \etal}{2010}]{moster10}
Moster, B. P., Maccio, A. V., Somerville, R. S., Johansson, P. H., Naab, T. 2010, MNRAS, 403, 1009M

\bibitem[\protect\citeauthoryear{Planck Collaboration XI}{2013}]{planck13}
Planck Collaboration, 2013, A\&A, 557, 52

\bibitem[\protect\citeauthoryear{Planck Collaboration XVI}{2013}]{planck13a}
Planck Collaboration, 2013, arXiv, 1303.507P


\bibitem[\protect\citeauthoryear{Putman \etal}{2012}]{putman12}
Putman, M. E., Peek, J. E. G., Joung, M. R. 2012, ARA\&A, 50, 491

\bibitem[\protect\citeauthoryear{Rassmussen \etal}{2009}]{rassmussen09}
Rassmussen, J., Sommer-Larsen, J., Pedersen, K. \etal 2009, ApJ, 697, 79

\bibitem[\protect\citeauthoryear{Silk}{1977}]{silk77}
Silk, J. 1977, ApJ, 211, 638

\bibitem[\protect\citeauthoryear{Sharma \etal}{2014}]{sharma14}
Sharma, M., Nath, B. B., Chattopadhyay, I., Shchekinov, Y. 2014, MNRAS, 441, 431

\bibitem[\protect\citeauthoryear{Sharma \etal}{2012}]{sharma12}
Sharma, P., McCourt, M., Parrish, I. J., Quataert, E. 2012, MNRAS, 427, 1219

\bibitem[\protect\citeauthoryear{Sheth \& Diaferio}{2001}]{sheth01a}
Sheth, R. K., Diaferio, A. 2001, MNRAS, 322, 901S

\bibitem[\protect\citeauthoryear{Sheth \& Tormen}{2001}]{sheth01}
Sheth, R. K, Mo, H. J., Tormen, G. 2001, MNRAS, 323, 1S


\bibitem[\protect\citeauthoryear{Sutherland \& Dopita}{1993}]{dopita93}
Sutherland, R. S., Dopita, M. A. 1993, ApJ, 88, 253S

\bibitem[\protect\citeauthoryear{Tumlinson \etal} {2011}]{tumlinson11}
Tumlinson, J., Thom, C., Werk, J. \etal 2011, Science, 334, 948

\bibitem[\protect\citeauthoryear{Walker \etal} {2014}]{walker14}
Walker, S. A., Bagchi, J., Fabian, A. C. 2014, arXiv:1411.1930v1 

\bibitem[\protect\citeauthoryear{Werk \etal} {2014}]{werk14}
Werk, J. K., Prochaska, J. X., Tumlinson, J. \etal 2014, arXiv:1403.0947

\bibitem[\protect\citeauthoryear{White \& Rees}{1978}]{white78}
White, S. D. M., Rees, M. J. 1978, MNRAS, 183, 341

\bibitem[\protect\citeauthoryear{White \& Frenk}{1991}]{white91}
White, S. D. M., Frenk, C. S. 1991, ApJ, 379, 52
 
\end{thebibliography}
\end{document}